%% file: knew1.tex
\documentstyle[preprint,prl,aps]{revtex}
\tighten
\begin{document}
%\preprint{\today}
\draft
%
%%%%%%%%%%%%%%%%%%%%%%%%%%%%%%%%% TITLE PAGE
%
\title{Quantum nondemolition measurements on two-level atomic systems 
and temporal Bell inequalities}
\author{Tommaso Calarco${}^{1}$ and Roberto Onofrio${}^{2}$} 
\address{${}^1$Dipartimento di Fisica, Universit\`a di Ferrara, and 
INFN, Sezione di Ferrara, \\ 
Via Paradiso 12, Ferrara, Italy 44100\\
${}^2$Dipartimento di Fisica ``G. Galilei'', 
Universit\`a di Padova and INFN, Sezione di Padova, \\ 
Via Marzolo 8, Padova, Italy 35131}
%\date{\today}
\maketitle
\begin{abstract}
The evolution of a two-level system subjected to stimulated transitions 
which is undergoing a sequence of measurements of the level occupation 
probability is evaluated. Its time correlation 
function is compared to the one obtained through the pure Schr\"odinger 
evolution. Systems of this kind have been recently proposed for 
testing the quantum mechanical predictions against those of macrorealistic 
theories, by means of temporal Bell inequalities.
The classical requirement of noninvasivity, needed to define correlation 
functions in the realistic case, finds
a quantum counterpart in the quantum nondemolition condition.
The consequences on the observability of quantum mechanically predicted
violations to temporal Bell inequalities are drawn and compared to the 
already dealt case of the rf-SQUID dynamics. 
\end{abstract}
%
%%%%%%%%%%%%%%%%%%%%%%%%%%%%%%%%% PACS NUMBERS
%
%
%%%%%%%%%%%%%%%%%%%%%%%%%%%%%%%%% PAPER CONTENT
%
%\narrowtext
The validity of quantum mechanics at the macroscopic 
level is still an open question crucial to understand why a 
particular limit of it, {\it classical mechanics}, work so well 
in a wide variety of situations visible to our eyes. 
Leggett and Garg have challenged this question by proposing laboratory tests 
aimed at comparing, in a macroscopic domain, the predictions of  a set of 
theories incorporating realism and noninvasivity, two properties manifestly not 
shared by quantum mechanics, and quantum mechanics itself \cite{LEGG}.
In analogy to the well-known {\it spatial} Bell inequalities \cite{BELL}, 
already tested \cite{ASPECT} and making light on the ultimate contrast of 
quantum mechanics with locality at the microscopic level,
Leggett and Garg have shown that certain relations among the correlation 
probabilities -~called {\it temporal} Bell inequalities~- 
which holds in realistic theories, are instead violated, 
with a proper choice of the measurement times, by  
the coherent evolution of the state dictated by quantum mechanics.   
The ingredients of temporal Bell inequalities, regardless of 
the concrete scheme used, are different-time correlation probabilities 
between subsequent measurements of a two-valued (dichotomic) observable.
However, the quantum mechanical predictions discussed so far do
not consider the effect of the Heisenberg principle on consecutive measurements 
of the same observable of the monitored system. In this paper we discuss this 
effect in the exactly solvable case of two-level systems, 
which have been recently proposed to experimentally test 
temporal Bell inequalities.
The first proposal  is based upon three two-level systems coupled through 
optical pulses \cite{PAZ}, one of which is monitored and the other two are 
treated as nondissipative memories which register the state of the first one at 
given times. 
The second proposal is based upon a Rydberg atom interacting with 
a single quantized mode of a superconducting resonant cavity \cite{SANTOS}. 
While Leggett and Garg claim that their proposed experiment gives insights 
on the validity of quantum mechanics at the macroscopic level \cite{LEGG}, 
the proposal in \cite{PAZ} is in a purely microscopic framework and the 
one in \cite{SANTOS} is located in between, an atomic system being involved, 
although in a large quantum number state, and interacting with a single 
{\it mesoscopic} mode of a QED cavity. 
In all these cases it turns out that the concept of quantum nondemolition 
(QND) measurements \cite{CAVES,BRAG} 
and its refinement to nearly QND measurements 
play a key role for understanding if violations to temporal Bell 
inequalities can be observed when somebody looks at them.

Temporal Bell inequalities are based upon different-time correlation functions,
calculable either in a classical (realistic) or in a quantum context.
The different-time correlation function for a generic observable 
$Q(t)$ can be written as \cite{PAZ} 
\begin{equation}
\label{defk}
K(t_1,t_2)\stackrel{\rm def}{\equiv}\int{{\cal D}[Q(t)]P[Q(t)]Q(t_1)Q(t_2)}
\end{equation}
where the information about the dynamics of the system is expressed through 
the probability functional $P[Q(t)]$, which selects the $Q(t)$ allowed by the 
dynamical evolution, possibly including the effect of the measurement.
This last can be easily taken into account by means of the concept of 
projection of the state \cite{vN}. 
Indeed, a quantum observable can be written 
in terms of its eigenvalues $q\in {\rm Sp}(\hat{Q})$, and the related 
projectors $\hat{P}_q$ (such that $\hat{P}_q^2=\hat{P}_q$) as 
$Q(t)=\sum_q q \hat{P}_q$
which implies 
\begin{equation}
\label{defq}
Q(t)\stackrel{\rm def}{\equiv}\frac{\langle\psi(t)|\hat{Q}|\psi(t)\rangle}
{\langle\psi(t)|\psi(t)\rangle}=
\sum_q q\frac{\langle\psi(t)|\hat{P}_q|\psi(t)\rangle}
{\langle\psi(t)|\psi(t)\rangle}.
\end{equation}
Before a measurement performed at time $\tilde{t}$,
\begin{equation}
Q(\tilde{t}^-)=\sum_q 
q\frac{\langle\psi(\tilde{t}^-)|\hat{P}_q|\psi(\tilde{t}^-)\rangle}
{\langle\psi(\tilde{t}^-)|\psi(\tilde{t}^-)\rangle};
\end{equation}
and, if the measurement result is ${\tilde{q}}$, after it we get
\begin{equation}
\label{collapse}
|\psi(\tilde{t}^+)\rangle=\hat{P}_{\tilde{q}}|\psi(\tilde{t}^-)\rangle
\end{equation}
and therefore 
\begin{equation}
\label{qmeas}
Q_{\tilde{q}}(\tilde{t}^+)={\tilde{q}}
\frac{\langle\psi(\tilde{t}^-)|\hat{P}_{\tilde{q}}|\psi(\tilde{t}^-)\rangle}
{\langle\psi(\tilde{t}^-)|\psi(\tilde{t}^-)\rangle}.
\end{equation}
Let us suppose, by starting from the state 
$|\psi(t_0)\rangle$, to measure the observable $\hat{Q}$ 
at $N$ instants of time $t_1<t_2<\cdots <t_N$ with outcomes 
$\{q_i\}_{1\leq i\leq N}$. Thus the successive evolutions 
of the state can be recursively written as
\begin{eqnarray}
\label{recurs}
|\psi(t_i^-)\rangle & = & 
U(t_i-t_{i-1})|\psi(t_{i-1}^+)\rangle;\nonumber\\
|\psi(t_i^+)\rangle & \stackrel{(\protect\ref{collapse})}{=} & 
\hat{P}_{q_i} U(t_i-t_{i-1})|\psi(t_{i-1}^+)\rangle.
\end{eqnarray}
The time-dependent expectation value $Q_{\{q_i\}}(t)$, including the effect of 
the $N$ measurements, can be calculated through Eqs.~(\ref{qmeas}), 
(\ref{recurs}).
The different times product of $Q$'s  required to define correlation 
functions as the (\ref{defk}), is written as
\begin{equation}
\label{qmqm}
Q_{\{q_i\}}(t_1)Q_{\{q_i\}}(t_2)\cdots Q_{\{q_i\}}(t_N)  = 
q_1q_2\cdots q_N\langle\psi(t_N^+)|\psi(t_N^+)\rangle.
\end{equation}
The $N$-times correlation function in the presence of 
measurements (in the case dealt here, $N=2$) can be evaluated by summing on the 
$Q_{\{q_i\}}(t)$ rather than on the $Q(t)$. 
The presence of measured trajectories $Q_{\{q_i\}}(t)$ replacing the
$Q(t)$ is not the only effect of the measurement process in Eqn.~(\ref{defk}).
The trajectory weigth $P[Q(t)]$ is also affected, since 
the need to measure in an actual experiment these 
correlations as joint probabilities for finding the system in {\em definite} 
states \cite{LEGG}, ``requires the obtained data to be purged by throwing 
away all the information coming from channels in which the state of the 
first memory changes'' \cite{PAZ}. 
The amount of off-line data processing and selection is 
therefore expressed as 
$\Delta=1-\langle\psi(t_1)|\hat{P}_{q_1}|\psi(t_1)\rangle$.
When $\Delta\not=0$, $Q_{\{q_i\}}(t)\not= Q(t)$ for $t>t_1$, 
the only exception in which $Q_{\{q_i\}}(t)= Q(t)$ is the impulsive QND case, 
when the system at the measurement time is already in the observed eigenstate.
More in general ideal QND stroboscopic measurements are obtained if they are 
performed each time interval corresponding to a complete 
reconstruction of the state (a complete revival of the wavefunction 
after the collapse induced by the measurement), as discussed in \cite{MEOP} 
for the case of position measurements on a generic nonlinear system. 
On the other hand, if $\Delta\not= 0$,the probability functional 
should select the $Q_{\{q_i\}}(t)$ for which $1-\Delta\geq\varepsilon$, where 
$\varepsilon$ is a distinguishability threshold which expresses the degree 
of reliability for the measurement to indicate a definite state of 
the system. This selection leads to the {\em selective} correlation
function $K_\varepsilon(t_1,t_2)$ which in the limit of complete selection 
becomes null. It follows that $K_\varepsilon(t_1,t_2)$ cannot violate temporal 
Bell inequalities if ideal QND measurements are required. Violations could 
be found for $\varepsilon=0$; however in this case the correlation function 
does not allow to distinguish the two eigenstates. We are looking for an 
intermediate regime in which a well-defined selective correlation function 
will show detectable violations to temporal Bell inequalities.
 
For a system with two energy levels $|+\rangle$ and $|-\rangle$, $\hat{Q}=\hat{P}_+-\hat{P}_-=|+\rangle\langle+|-|-\rangle\langle-|$ and
\begin{equation}
Q(t)=|\langle +|\psi(t)\rangle |^2-|\langle -|\psi(t)\rangle |^2
\end{equation}
which holds both for a spin-$\frac 12$ system \cite{PAZ} and for an atom 
coupled to a single mode of a resonant cavity \cite{SANTOS}, provided that 
$|+\rangle$ and $|-\rangle$ stand for the excited and the ground 
state respectively. 
If the system is harmonically oscillating between the two states, 
the matrix elements involved in the evaluation of Eqn.~(\ref{qmqm}) are
\begin{equation}
\label{qexpand}
\begin{array}{lll}
\langle +|U(t-t_0)|+\rangle=\cos \omega(t-t_0)& \qquad &
\langle +|U(t-t_0)|-\rangle=-\sin \omega(t-t_0)\\
\langle -|U(t-t_0)|+\rangle=\sin \omega(t-t_0)& \qquad &
\langle -|U(t-t_0)|-\rangle=\cos \omega(t-t_0).
\end{array}
\end{equation}
For the spin system considered in \cite{PAZ}, $\omega$ is the frequency of Rabi 
oscillations $\Omega_R$, whereas for the atom-cavity system \cite{SANTOS} the
Jaynes-Cummings evolution  gives $\omega=\Omega_R\sqrt{n+1}$ (where $n$
is the principal quantum number of the Rydberg excited state, for 
an experimental demonstration on single atoms see \cite{REMPE}).\\
The initial state of the system $|\psi(t_0)\rangle =c_+(t_0)|+\rangle 
+c_-(t_0)|-\rangle$ (with $|c_+(t_0)|^2+|c_-(t_0)|^2=1$) can  be parametrized 
with $c_+(t_0)=\cos \omega(t-t')$, $c_-(t_0)=\sin \omega(t-t')$. From 
Eqn.~(\ref{qexpand}) then follows $Q(t)=\cos 2\omega(t-t')$, where 
$t'$ is the only free parameter (representing some arbitrary instant at which 
the system was in $|+\rangle$) on which one should integrate for evaluating the correlation function:
\begin{equation}
K(t_1,t_2)=\frac{\omega}{\pi}\int_0^{\frac{2\pi}{\omega}}
\cos 2\omega(t'-t_1)\cos 2\omega(t'-t_2)dt'=\cos 2\omega(t_1-t_2),
\end{equation}
which depends only upon the time difference $t_2-t_1$. 
The distinguishability condition is expressed as
\begin{equation}
\label{disting}
\langle\psi(t_1)|\hat{P}_{q_1}|\psi(t_1)\rangle\geq\varepsilon.
\end{equation}
The propagator including the effect of the measurement is calculated 
by summing over $n_1$ and $n_2$ and integrating on $t_0$ under the condition
(\ref{disting}), obtaining a factorized form
\begin{equation}
\label{defkeps}
K_\varepsilon(t_1,t_2)=
\frac{1}{\pi}\left[2\sqrt{\varepsilon(1-\varepsilon)}+
\arccos(2\varepsilon-1)\right]
K(t_2-t_1)\stackrel{\rm def}{\equiv}A_\varepsilon K(t_2-t_1),
\end{equation}
which shows the correct limits $K_0(t_1,t_2)=K(t_1,t_2)$ and $K_1(t_1,t_2)=0$. 

A generic temporal Bell inequality involves a combination $\Delta K$ of 
two-time correlation functions $K(t_i,t_j)$ with some coefficients $\kappa_{ij}$
and an upper bound $B$:
\begin{equation}
\Delta K=\sum_{i\not= j=1}^N\kappa_{ij}K(t_i,t_j)\leq B.
\end{equation}
which can be violated for some values of $\{(t_i,t_j)\}_{i\not= j=1,\ldots,N}$
by the quantum mechanical predictions, with a maximum 
$\Delta K_{\rm max}\stackrel{\rm def}{\equiv}\max_{\{(t_i,t_j)\}}\Delta K>B$. 
For the system described in \cite{PAZ}, $B=2$ and 
\begin{equation}
\Delta K=|K(t_1,t_2)+K(t_2,t_3)+K(t_3,t_4)-K(t_1,t_4)|\qquad\Delta K_{\rm 
max}=2\sqrt{2}
\end{equation}
whereas in \cite{SANTOS} $B=1$ and
\begin{eqnarray}
\Delta K_- & = & -K(t_1,t_2)-K(t_2,t_3)-K(t_1,t_3)\qquad\Delta K_{\rm max}=
3/2
\nonumber\\
\Delta K_+ & = & -K(t_1,t_3)+K(t_1,t_2)+K(t_2,t_3)\qquad\Delta K_{\rm max}=
3/2.
\end{eqnarray}
$\Delta K_\pm$ can be simplified by introducing the so-called 
stationarity assumption \cite{SANTOS}, namely $t_2-t_1=t_3-t_2=t$, 
corresponding to stroboscopic (equally spaced) measurements, such that $\Delta
K$ depends only on $t$. For instance, Fig.~1 shows on the same scale the 
behavior of $\Delta K_-(t)$ and of $Q(t)$. 
The upper bound for the fulfilment of the corresponding temporal 
Bell inequality is also depicted. 
The impossibility of simultaneously having ideal QND 
measurements of the occupation number, corresponding to a complete revival 
of the initial state, and maximal 
violations to temporal Bell inequalities is evidenced. 
To refine in a quantitative way the possibility of coexistence of unoptimal 
violations to the temporal Bell inequalities and quasi QND measurements 
one can use the distinguishability level $\varepsilon$. 
The measurement effect is represented, as in 
Eqn.~(\ref{defkeps}), by a factor $A_\varepsilon$ multiplying $\Delta K$. 
It is worth noting that $A_\varepsilon$ is independent upon time and thus 
leaves unchanged the correlation times for which $\Delta K$ is maximal. 
The maximal violation in percentage under the effect of the measurements is
then expressed as 
$\Delta B_{\rm max}=(A_\varepsilon\Delta K_{\rm max}-B)/B$.
Fig.~2 shows $\Delta B_{\rm max}$
versus the distinguishability level $\varepsilon$.
As already observed in \cite{SANTOS} without measurement effects, we 
confirm here that even in presence of measurements, 
by assuming the same distinguishability level,  the proposed  
inequalities are more violated than in \cite{PAZ}. 
As expected, the violations disappear 
for nearly QND measurements, but are present for $\varepsilon\leq 0.693$.
So the proposed systems could be used for testing 
the predictions of quantum mechanics against those of a realistic theory, 
as claimed in \cite{PAZ,SANTOS}, only if one requires a reliability,
for the detection of distinct states, not greater than 70\%.

The examined experiments should be performed by looking at the 
correlation functions of an already-dichotomic variable, the state 
in two-level systems; the one discussed in \cite{LEGG,TESCHE} 
deals with the reduction of the spectrum of a {\em continuous} observable, the 
magnetic flux $\phi$ trapped in a SQUID, into a dichotomic variable, its 
sign ${\hat{\phi}}/{|\hat{\phi}|}$. 
This proposal  has been already discussed in \cite{CALON,ONCAL} by also 
including the effect of the Heisenberg principle, showing that 
for selective measurements there is incompatibility between violation 
of the inequalities and distinguishability of the dichotomic variable. 
Work is in progress to repeat the same analysis developed here 
for the case of the rf-SQUIDs dynamics. In this case the projectors on states 
of definite sign are $P_\pm=\Theta(\pm\phi)$: from Eqn.~(\ref{defq}) then follows
\begin{equation}
Q(t)=\int_0^\infty|\psi(\phi)|^2d\phi-\int_{-\infty}^0|\psi(\phi)|^2d\phi.
\end{equation}
(if $\psi(\phi)$ is normalized to 1)
and $K_\varepsilon(t_1,t_2)$ can be calculated by Eqn.~(\ref{defkeps}), as we 
will report in a forthcoming paper \cite{CALON1}.

In conclusion, our result can be simply summarized as follows: the
observability of violations depends critically on the statistical criterion 
adopted for defining the resolution of distinct states. In other words,
violations to temporal Bell inequalities can be detected only in a {\em probabilistic} way, unlike the spatial case. This is due to the fact that the 
time-dependent correlation probabilities, unlike the space-dependent
ones, involve measurements on the {\em same} observable of the {\em same}
system, and therefore the quantum predictions are characterized by an 
uncertainty dictated by the Heisenberg principle, which makes ambiguous the 
definition of the violations themselves.
%%%%%%%%%%%%%%%%%%%%%%%%%%%%%%%%% ACKNOWLEDGMENTS
%\acknowledgments 
%
%%%%%%%%%%%%%%%%%%%%%%%%%%%%%%%%% REFERENCES LIST
%

%%%%%%%%%%%%%%%%%%%%%%%%%%%%%FIGURES
%
\begin{figure}[h]
\include{fig1}
\caption{Time evolution of the two-level system in \protect\cite{SANTOS}. The
time-dependent expectation value of the observable $Q(t)$ (thick line) and the
temporal Bell inequality parameter $\Delta K(t)$ (thin line) with its upper
bound $B$ (dotted line) are represented on the same scale. Ideal QND 
measurements correspond to time intervals integer multiples 
of $\pi/\omega$, for which the inequality is not violated; 
however, small violations are compatible
with quasi QND measurements around odd-integer multiples of $\pi/\omega$.}
\end{figure}
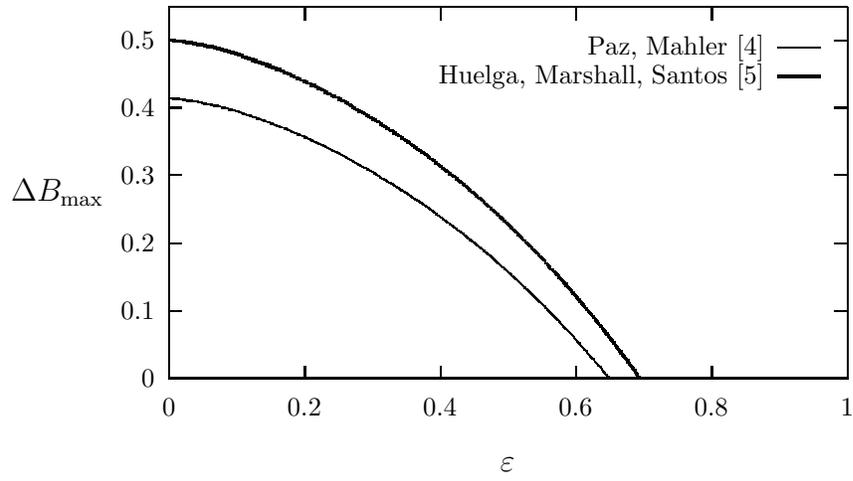
\begin{figure}[h]
\include{fig2}
\caption{Observability of violations to temporal Bell inequalities. 
The dependence of the maximal violation $\Delta B_{\rm max}$ upon the
distinguishability $\varepsilon$ is depicted for both the experiments 
proposed in \protect\cite{PAZ} and\protect\cite{SANTOS}. 
Violations disappear in the QND limit ($\varepsilon\to 1$), and are present 
only for $\varepsilon\leq 0.649$ \protect\cite{PAZ} and  
$\varepsilon\leq 0.693$ \protect\cite{SANTOS}.}
\end{figure}
\end{document}

%% file: fig1.tex
% GNUPLOT: LaTeX picture
\setlength{\unitlength}{0.240900pt}
\ifx\plotpoint\undefined\newsavebox{\plotpoint}\fi
\sbox{\plotpoint}{\rule[-0.200pt]{0.400pt}{0.400pt}}%
\begin{picture}(1349,720)(0,0)
\font\gnuplot=cmr10 at 10pt
\gnuplot
\sbox{\plotpoint}{\rule[-0.200pt]{0.400pt}{0.400pt}}%
\put(176.0,113.0){\rule[-0.200pt]{0.400pt}{140.686pt}}
\put(176.0,113.0){\rule[-0.200pt]{4.818pt}{0.400pt}}
\put(154,113){\makebox(0,0)[r]{-3}}
\put(1265.0,113.0){\rule[-0.200pt]{4.818pt}{0.400pt}}
\put(176.0,178.0){\rule[-0.200pt]{4.818pt}{0.400pt}}
\put(154,178){\makebox(0,0)[r]{-2.5}}
\put(1265.0,178.0){\rule[-0.200pt]{4.818pt}{0.400pt}}
\put(176.0,243.0){\rule[-0.200pt]{4.818pt}{0.400pt}}
\put(154,243){\makebox(0,0)[r]{-2}}
\put(1265.0,243.0){\rule[-0.200pt]{4.818pt}{0.400pt}}
\put(176.0,308.0){\rule[-0.200pt]{4.818pt}{0.400pt}}
\put(154,308){\makebox(0,0)[r]{-1.5}}
\put(1265.0,308.0){\rule[-0.200pt]{4.818pt}{0.400pt}}
\put(176.0,373.0){\rule[-0.200pt]{4.818pt}{0.400pt}}
\put(154,373){\makebox(0,0)[r]{-1}}
\put(1265.0,373.0){\rule[-0.200pt]{4.818pt}{0.400pt}}
\put(176.0,437.0){\rule[-0.200pt]{4.818pt}{0.400pt}}
\put(154,437){\makebox(0,0)[r]{-0.5}}
\put(1265.0,437.0){\rule[-0.200pt]{4.818pt}{0.400pt}}
\put(176.0,502.0){\rule[-0.200pt]{4.818pt}{0.400pt}}
\put(154,502){\makebox(0,0)[r]{0}}
\put(1265.0,502.0){\rule[-0.200pt]{4.818pt}{0.400pt}}
\put(176.0,567.0){\rule[-0.200pt]{4.818pt}{0.400pt}}
\put(154,567){\makebox(0,0)[r]{0.5}}
\put(1265.0,567.0){\rule[-0.200pt]{4.818pt}{0.400pt}}
\put(176.0,632.0){\rule[-0.200pt]{4.818pt}{0.400pt}}
\put(154,632){\makebox(0,0)[r]{1}}
\put(1265.0,632.0){\rule[-0.200pt]{4.818pt}{0.400pt}}
\put(176.0,697.0){\rule[-0.200pt]{4.818pt}{0.400pt}}
\put(154,697){\makebox(0,0)[r]{1.5}}
\put(1265.0,697.0){\rule[-0.200pt]{4.818pt}{0.400pt}}
\put(176.0,113.0){\rule[-0.200pt]{0.400pt}{4.818pt}}
\put(176,68){\makebox(0,0){0}}
\put(176.0,677.0){\rule[-0.200pt]{0.400pt}{4.818pt}}
\put(453.0,113.0){\rule[-0.200pt]{0.400pt}{4.818pt}}
\put(453,68){\makebox(0,0){$\pi$}}
\put(453.0,677.0){\rule[-0.200pt]{0.400pt}{4.818pt}}
\put(730.0,113.0){\rule[-0.200pt]{0.400pt}{4.818pt}}
\put(730,68){\makebox(0,0){$2\pi$}}
\put(730.0,677.0){\rule[-0.200pt]{0.400pt}{4.818pt}}
\put(1008.0,113.0){\rule[-0.200pt]{0.400pt}{4.818pt}}
\put(1008,68){\makebox(0,0){$3\pi$}}
\put(1008.0,677.0){\rule[-0.200pt]{0.400pt}{4.818pt}}
\put(1285.0,113.0){\rule[-0.200pt]{0.400pt}{4.818pt}}
\put(1285,68){\makebox(0,0){$4\pi$}}
\put(1285.0,677.0){\rule[-0.200pt]{0.400pt}{4.818pt}}
\put(176.0,113.0){\rule[-0.200pt]{267.158pt}{0.400pt}}
\put(1285.0,113.0){\rule[-0.200pt]{0.400pt}{140.686pt}}
\put(176.0,697.0){\rule[-0.200pt]{267.158pt}{0.400pt}}
\put(730,-22){\makebox(0,0){$\omega t$}}
\put(176.0,113.0){\rule[-0.200pt]{0.400pt}{140.686pt}}
\sbox{\plotpoint}{\rule[-0.400pt]{0.800pt}{0.800pt}}%
\put(1059,308){\makebox(0,0)[r]{$Q(t)$}}
\put(1081.0,308.0){\rule[-0.400pt]{15.899pt}{0.800pt}}
\put(176,632){\usebox{\plotpoint}}
\put(176,629.84){\rule{2.650pt}{0.800pt}}
\multiput(176.00,630.34)(5.500,-1.000){2}{\rule{1.325pt}{0.800pt}}
\put(187,627.84){\rule{2.650pt}{0.800pt}}
\multiput(187.00,629.34)(5.500,-3.000){2}{\rule{1.325pt}{0.800pt}}
\multiput(198.00,626.06)(1.600,-0.560){3}{\rule{2.120pt}{0.135pt}}
\multiput(198.00,626.34)(7.600,-5.000){2}{\rule{1.060pt}{0.800pt}}
\multiput(210.00,621.08)(0.825,-0.526){7}{\rule{1.457pt}{0.127pt}}
\multiput(210.00,621.34)(7.976,-7.000){2}{\rule{0.729pt}{0.800pt}}
\multiput(221.00,614.08)(0.611,-0.516){11}{\rule{1.178pt}{0.124pt}}
\multiput(221.00,614.34)(8.555,-9.000){2}{\rule{0.589pt}{0.800pt}}
\multiput(232.00,605.08)(0.489,-0.512){15}{\rule{1.000pt}{0.123pt}}
\multiput(232.00,605.34)(8.924,-11.000){2}{\rule{0.500pt}{0.800pt}}
\multiput(244.40,591.55)(0.512,-0.539){15}{\rule{0.123pt}{1.073pt}}
\multiput(241.34,593.77)(11.000,-9.774){2}{\rule{0.800pt}{0.536pt}}
\multiput(255.41,579.57)(0.511,-0.536){17}{\rule{0.123pt}{1.067pt}}
\multiput(252.34,581.79)(12.000,-10.786){2}{\rule{0.800pt}{0.533pt}}
\multiput(267.40,565.64)(0.512,-0.689){15}{\rule{0.123pt}{1.291pt}}
\multiput(264.34,568.32)(11.000,-12.321){2}{\rule{0.800pt}{0.645pt}}
\multiput(278.40,550.64)(0.512,-0.689){15}{\rule{0.123pt}{1.291pt}}
\multiput(275.34,553.32)(11.000,-12.321){2}{\rule{0.800pt}{0.645pt}}
\multiput(289.40,535.34)(0.512,-0.739){15}{\rule{0.123pt}{1.364pt}}
\multiput(286.34,538.17)(11.000,-13.170){2}{\rule{0.800pt}{0.682pt}}
\multiput(300.40,519.34)(0.512,-0.739){15}{\rule{0.123pt}{1.364pt}}
\multiput(297.34,522.17)(11.000,-13.170){2}{\rule{0.800pt}{0.682pt}}
\multiput(311.41,503.47)(0.511,-0.717){17}{\rule{0.123pt}{1.333pt}}
\multiput(308.34,506.23)(12.000,-14.233){2}{\rule{0.800pt}{0.667pt}}
\multiput(323.40,486.34)(0.512,-0.739){15}{\rule{0.123pt}{1.364pt}}
\multiput(320.34,489.17)(11.000,-13.170){2}{\rule{0.800pt}{0.682pt}}
\multiput(334.40,470.34)(0.512,-0.739){15}{\rule{0.123pt}{1.364pt}}
\multiput(331.34,473.17)(11.000,-13.170){2}{\rule{0.800pt}{0.682pt}}
\multiput(345.40,454.64)(0.512,-0.689){15}{\rule{0.123pt}{1.291pt}}
\multiput(342.34,457.32)(11.000,-12.321){2}{\rule{0.800pt}{0.645pt}}
\multiput(356.40,439.64)(0.512,-0.689){15}{\rule{0.123pt}{1.291pt}}
\multiput(353.34,442.32)(11.000,-12.321){2}{\rule{0.800pt}{0.645pt}}
\multiput(367.41,425.57)(0.511,-0.536){17}{\rule{0.123pt}{1.067pt}}
\multiput(364.34,427.79)(12.000,-10.786){2}{\rule{0.800pt}{0.533pt}}
\multiput(378.00,415.08)(0.489,-0.512){15}{\rule{1.000pt}{0.123pt}}
\multiput(378.00,415.34)(8.924,-11.000){2}{\rule{0.500pt}{0.800pt}}
\multiput(389.00,404.08)(0.489,-0.512){15}{\rule{1.000pt}{0.123pt}}
\multiput(389.00,404.34)(8.924,-11.000){2}{\rule{0.500pt}{0.800pt}}
\multiput(400.00,393.08)(0.700,-0.520){9}{\rule{1.300pt}{0.125pt}}
\multiput(400.00,393.34)(8.302,-8.000){2}{\rule{0.650pt}{0.800pt}}
\multiput(411.00,385.08)(0.825,-0.526){7}{\rule{1.457pt}{0.127pt}}
\multiput(411.00,385.34)(7.976,-7.000){2}{\rule{0.729pt}{0.800pt}}
\put(422,376.34){\rule{2.600pt}{0.800pt}}
\multiput(422.00,378.34)(6.604,-4.000){2}{\rule{1.300pt}{0.800pt}}
\put(434,372.84){\rule{2.650pt}{0.800pt}}
\multiput(434.00,374.34)(5.500,-3.000){2}{\rule{1.325pt}{0.800pt}}
\put(456,371.84){\rule{2.650pt}{0.800pt}}
\multiput(456.00,371.34)(5.500,1.000){2}{\rule{1.325pt}{0.800pt}}
\put(467,374.34){\rule{2.400pt}{0.800pt}}
\multiput(467.00,372.34)(6.019,4.000){2}{\rule{1.200pt}{0.800pt}}
\multiput(478.00,379.38)(1.600,0.560){3}{\rule{2.120pt}{0.135pt}}
\multiput(478.00,376.34)(7.600,5.000){2}{\rule{1.060pt}{0.800pt}}
\multiput(490.00,384.40)(0.700,0.520){9}{\rule{1.300pt}{0.125pt}}
\multiput(490.00,381.34)(8.302,8.000){2}{\rule{0.650pt}{0.800pt}}
\multiput(501.00,392.40)(0.611,0.516){11}{\rule{1.178pt}{0.124pt}}
\multiput(501.00,389.34)(8.555,9.000){2}{\rule{0.589pt}{0.800pt}}
\multiput(512.00,401.40)(0.489,0.512){15}{\rule{1.000pt}{0.123pt}}
\multiput(512.00,398.34)(8.924,11.000){2}{\rule{0.500pt}{0.800pt}}
\multiput(524.40,411.00)(0.512,0.589){15}{\rule{0.123pt}{1.145pt}}
\multiput(521.34,411.00)(11.000,10.623){2}{\rule{0.800pt}{0.573pt}}
\multiput(535.41,424.00)(0.511,0.536){17}{\rule{0.123pt}{1.067pt}}
\multiput(532.34,424.00)(12.000,10.786){2}{\rule{0.800pt}{0.533pt}}
\multiput(547.40,437.00)(0.512,0.689){15}{\rule{0.123pt}{1.291pt}}
\multiput(544.34,437.00)(11.000,12.321){2}{\rule{0.800pt}{0.645pt}}
\multiput(558.40,452.00)(0.512,0.739){15}{\rule{0.123pt}{1.364pt}}
\multiput(555.34,452.00)(11.000,13.170){2}{\rule{0.800pt}{0.682pt}}
\multiput(569.40,468.00)(0.512,0.739){15}{\rule{0.123pt}{1.364pt}}
\multiput(566.34,468.00)(11.000,13.170){2}{\rule{0.800pt}{0.682pt}}
\multiput(580.40,484.00)(0.512,0.739){15}{\rule{0.123pt}{1.364pt}}
\multiput(577.34,484.00)(11.000,13.170){2}{\rule{0.800pt}{0.682pt}}
\multiput(591.41,500.00)(0.511,0.717){17}{\rule{0.123pt}{1.333pt}}
\multiput(588.34,500.00)(12.000,14.233){2}{\rule{0.800pt}{0.667pt}}
\multiput(603.40,517.00)(0.512,0.739){15}{\rule{0.123pt}{1.364pt}}
\multiput(600.34,517.00)(11.000,13.170){2}{\rule{0.800pt}{0.682pt}}
\multiput(614.40,533.00)(0.512,0.739){15}{\rule{0.123pt}{1.364pt}}
\multiput(611.34,533.00)(11.000,13.170){2}{\rule{0.800pt}{0.682pt}}
\multiput(625.40,549.00)(0.512,0.689){15}{\rule{0.123pt}{1.291pt}}
\multiput(622.34,549.00)(11.000,12.321){2}{\rule{0.800pt}{0.645pt}}
\multiput(636.40,564.00)(0.512,0.639){15}{\rule{0.123pt}{1.218pt}}
\multiput(633.34,564.00)(11.000,11.472){2}{\rule{0.800pt}{0.609pt}}
\multiput(646.00,579.41)(0.491,0.511){17}{\rule{1.000pt}{0.123pt}}
\multiput(646.00,576.34)(9.924,12.000){2}{\rule{0.500pt}{0.800pt}}
\multiput(659.40,590.00)(0.512,0.539){15}{\rule{0.123pt}{1.073pt}}
\multiput(656.34,590.00)(11.000,9.774){2}{\rule{0.800pt}{0.536pt}}
\multiput(669.00,603.40)(0.543,0.514){13}{\rule{1.080pt}{0.124pt}}
\multiput(669.00,600.34)(8.758,10.000){2}{\rule{0.540pt}{0.800pt}}
\multiput(680.00,613.40)(0.700,0.520){9}{\rule{1.300pt}{0.125pt}}
\multiput(680.00,610.34)(8.302,8.000){2}{\rule{0.650pt}{0.800pt}}
\multiput(691.00,621.39)(1.020,0.536){5}{\rule{1.667pt}{0.129pt}}
\multiput(691.00,618.34)(7.541,6.000){2}{\rule{0.833pt}{0.800pt}}
\put(702,626.34){\rule{2.600pt}{0.800pt}}
\multiput(702.00,624.34)(6.604,4.000){2}{\rule{1.300pt}{0.800pt}}
\put(714,629.34){\rule{2.650pt}{0.800pt}}
\multiput(714.00,628.34)(5.500,2.000){2}{\rule{1.325pt}{0.800pt}}
\put(445.0,373.0){\rule[-0.400pt]{2.650pt}{0.800pt}}
\put(736,629.34){\rule{2.650pt}{0.800pt}}
\multiput(736.00,630.34)(5.500,-2.000){2}{\rule{1.325pt}{0.800pt}}
\put(747,626.34){\rule{2.600pt}{0.800pt}}
\multiput(747.00,628.34)(6.604,-4.000){2}{\rule{1.300pt}{0.800pt}}
\multiput(759.00,624.07)(1.020,-0.536){5}{\rule{1.667pt}{0.129pt}}
\multiput(759.00,624.34)(7.541,-6.000){2}{\rule{0.833pt}{0.800pt}}
\multiput(770.00,618.08)(0.700,-0.520){9}{\rule{1.300pt}{0.125pt}}
\multiput(770.00,618.34)(8.302,-8.000){2}{\rule{0.650pt}{0.800pt}}
\multiput(781.00,610.08)(0.543,-0.514){13}{\rule{1.080pt}{0.124pt}}
\multiput(781.00,610.34)(8.758,-10.000){2}{\rule{0.540pt}{0.800pt}}
\multiput(793.40,597.55)(0.512,-0.539){15}{\rule{0.123pt}{1.073pt}}
\multiput(790.34,599.77)(11.000,-9.774){2}{\rule{0.800pt}{0.536pt}}
\multiput(803.00,588.08)(0.491,-0.511){17}{\rule{1.000pt}{0.123pt}}
\multiput(803.00,588.34)(9.924,-12.000){2}{\rule{0.500pt}{0.800pt}}
\multiput(816.40,572.94)(0.512,-0.639){15}{\rule{0.123pt}{1.218pt}}
\multiput(813.34,575.47)(11.000,-11.472){2}{\rule{0.800pt}{0.609pt}}
\multiput(827.40,558.64)(0.512,-0.689){15}{\rule{0.123pt}{1.291pt}}
\multiput(824.34,561.32)(11.000,-12.321){2}{\rule{0.800pt}{0.645pt}}
\multiput(838.40,543.34)(0.512,-0.739){15}{\rule{0.123pt}{1.364pt}}
\multiput(835.34,546.17)(11.000,-13.170){2}{\rule{0.800pt}{0.682pt}}
\multiput(849.40,527.34)(0.512,-0.739){15}{\rule{0.123pt}{1.364pt}}
\multiput(846.34,530.17)(11.000,-13.170){2}{\rule{0.800pt}{0.682pt}}
\multiput(860.41,511.47)(0.511,-0.717){17}{\rule{0.123pt}{1.333pt}}
\multiput(857.34,514.23)(12.000,-14.233){2}{\rule{0.800pt}{0.667pt}}
\multiput(872.40,494.34)(0.512,-0.739){15}{\rule{0.123pt}{1.364pt}}
\multiput(869.34,497.17)(11.000,-13.170){2}{\rule{0.800pt}{0.682pt}}
\multiput(883.40,478.34)(0.512,-0.739){15}{\rule{0.123pt}{1.364pt}}
\multiput(880.34,481.17)(11.000,-13.170){2}{\rule{0.800pt}{0.682pt}}
\multiput(894.40,462.34)(0.512,-0.739){15}{\rule{0.123pt}{1.364pt}}
\multiput(891.34,465.17)(11.000,-13.170){2}{\rule{0.800pt}{0.682pt}}
\multiput(905.40,446.64)(0.512,-0.689){15}{\rule{0.123pt}{1.291pt}}
\multiput(902.34,449.32)(11.000,-12.321){2}{\rule{0.800pt}{0.645pt}}
\multiput(916.41,432.57)(0.511,-0.536){17}{\rule{0.123pt}{1.067pt}}
\multiput(913.34,434.79)(12.000,-10.786){2}{\rule{0.800pt}{0.533pt}}
\multiput(928.40,419.25)(0.512,-0.589){15}{\rule{0.123pt}{1.145pt}}
\multiput(925.34,421.62)(11.000,-10.623){2}{\rule{0.800pt}{0.573pt}}
\multiput(938.00,409.08)(0.489,-0.512){15}{\rule{1.000pt}{0.123pt}}
\multiput(938.00,409.34)(8.924,-11.000){2}{\rule{0.500pt}{0.800pt}}
\multiput(949.00,398.08)(0.611,-0.516){11}{\rule{1.178pt}{0.124pt}}
\multiput(949.00,398.34)(8.555,-9.000){2}{\rule{0.589pt}{0.800pt}}
\multiput(960.00,389.08)(0.700,-0.520){9}{\rule{1.300pt}{0.125pt}}
\multiput(960.00,389.34)(8.302,-8.000){2}{\rule{0.650pt}{0.800pt}}
\multiput(971.00,381.06)(1.600,-0.560){3}{\rule{2.120pt}{0.135pt}}
\multiput(971.00,381.34)(7.600,-5.000){2}{\rule{1.060pt}{0.800pt}}
\put(983,374.34){\rule{2.400pt}{0.800pt}}
\multiput(983.00,376.34)(6.019,-4.000){2}{\rule{1.200pt}{0.800pt}}
\put(994,371.84){\rule{2.650pt}{0.800pt}}
\multiput(994.00,372.34)(5.500,-1.000){2}{\rule{1.325pt}{0.800pt}}
\put(725.0,632.0){\rule[-0.400pt]{2.650pt}{0.800pt}}
\put(1016,372.84){\rule{2.650pt}{0.800pt}}
\multiput(1016.00,371.34)(5.500,3.000){2}{\rule{1.325pt}{0.800pt}}
\put(1027,376.34){\rule{2.600pt}{0.800pt}}
\multiput(1027.00,374.34)(6.604,4.000){2}{\rule{1.300pt}{0.800pt}}
\multiput(1039.00,381.40)(0.825,0.526){7}{\rule{1.457pt}{0.127pt}}
\multiput(1039.00,378.34)(7.976,7.000){2}{\rule{0.729pt}{0.800pt}}
\multiput(1050.00,388.40)(0.700,0.520){9}{\rule{1.300pt}{0.125pt}}
\multiput(1050.00,385.34)(8.302,8.000){2}{\rule{0.650pt}{0.800pt}}
\multiput(1061.00,396.40)(0.489,0.512){15}{\rule{1.000pt}{0.123pt}}
\multiput(1061.00,393.34)(8.924,11.000){2}{\rule{0.500pt}{0.800pt}}
\multiput(1072.00,407.40)(0.489,0.512){15}{\rule{1.000pt}{0.123pt}}
\multiput(1072.00,404.34)(8.924,11.000){2}{\rule{0.500pt}{0.800pt}}
\multiput(1084.41,417.00)(0.511,0.536){17}{\rule{0.123pt}{1.067pt}}
\multiput(1081.34,417.00)(12.000,10.786){2}{\rule{0.800pt}{0.533pt}}
\multiput(1096.40,430.00)(0.512,0.689){15}{\rule{0.123pt}{1.291pt}}
\multiput(1093.34,430.00)(11.000,12.321){2}{\rule{0.800pt}{0.645pt}}
\multiput(1107.40,445.00)(0.512,0.689){15}{\rule{0.123pt}{1.291pt}}
\multiput(1104.34,445.00)(11.000,12.321){2}{\rule{0.800pt}{0.645pt}}
\multiput(1118.40,460.00)(0.512,0.739){15}{\rule{0.123pt}{1.364pt}}
\multiput(1115.34,460.00)(11.000,13.170){2}{\rule{0.800pt}{0.682pt}}
\multiput(1129.40,476.00)(0.512,0.739){15}{\rule{0.123pt}{1.364pt}}
\multiput(1126.34,476.00)(11.000,13.170){2}{\rule{0.800pt}{0.682pt}}
\multiput(1140.41,492.00)(0.511,0.717){17}{\rule{0.123pt}{1.333pt}}
\multiput(1137.34,492.00)(12.000,14.233){2}{\rule{0.800pt}{0.667pt}}
\multiput(1152.40,509.00)(0.512,0.739){15}{\rule{0.123pt}{1.364pt}}
\multiput(1149.34,509.00)(11.000,13.170){2}{\rule{0.800pt}{0.682pt}}
\multiput(1163.40,525.00)(0.512,0.739){15}{\rule{0.123pt}{1.364pt}}
\multiput(1160.34,525.00)(11.000,13.170){2}{\rule{0.800pt}{0.682pt}}
\multiput(1174.40,541.00)(0.512,0.689){15}{\rule{0.123pt}{1.291pt}}
\multiput(1171.34,541.00)(11.000,12.321){2}{\rule{0.800pt}{0.645pt}}
\multiput(1185.40,556.00)(0.512,0.689){15}{\rule{0.123pt}{1.291pt}}
\multiput(1182.34,556.00)(11.000,12.321){2}{\rule{0.800pt}{0.645pt}}
\multiput(1196.41,571.00)(0.511,0.536){17}{\rule{0.123pt}{1.067pt}}
\multiput(1193.34,571.00)(12.000,10.786){2}{\rule{0.800pt}{0.533pt}}
\multiput(1208.40,584.00)(0.512,0.539){15}{\rule{0.123pt}{1.073pt}}
\multiput(1205.34,584.00)(11.000,9.774){2}{\rule{0.800pt}{0.536pt}}
\multiput(1218.00,597.40)(0.489,0.512){15}{\rule{1.000pt}{0.123pt}}
\multiput(1218.00,594.34)(8.924,11.000){2}{\rule{0.500pt}{0.800pt}}
\multiput(1229.00,608.40)(0.611,0.516){11}{\rule{1.178pt}{0.124pt}}
\multiput(1229.00,605.34)(8.555,9.000){2}{\rule{0.589pt}{0.800pt}}
\multiput(1240.00,617.40)(0.825,0.526){7}{\rule{1.457pt}{0.127pt}}
\multiput(1240.00,614.34)(7.976,7.000){2}{\rule{0.729pt}{0.800pt}}
\multiput(1251.00,624.38)(1.600,0.560){3}{\rule{2.120pt}{0.135pt}}
\multiput(1251.00,621.34)(7.600,5.000){2}{\rule{1.060pt}{0.800pt}}
\put(1263,627.84){\rule{2.650pt}{0.800pt}}
\multiput(1263.00,626.34)(5.500,3.000){2}{\rule{1.325pt}{0.800pt}}
\put(1274,629.84){\rule{2.650pt}{0.800pt}}
\multiput(1274.00,629.34)(5.500,1.000){2}{\rule{1.325pt}{0.800pt}}
\put(1005.0,373.0){\rule[-0.400pt]{2.650pt}{0.800pt}}
\sbox{\plotpoint}{\rule[-0.200pt]{0.400pt}{0.400pt}}%
\put(1059,263){\makebox(0,0)[r]{$\Delta K_-$}}
\put(1081.0,263.0){\rule[-0.200pt]{15.899pt}{0.400pt}}
\put(176,113){\usebox{\plotpoint}}
\multiput(176.00,113.59)(0.943,0.482){9}{\rule{0.833pt}{0.116pt}}
\multiput(176.00,112.17)(9.270,6.000){2}{\rule{0.417pt}{0.400pt}}
\multiput(187.58,119.00)(0.492,0.873){19}{\rule{0.118pt}{0.791pt}}
\multiput(186.17,119.00)(11.000,17.358){2}{\rule{0.400pt}{0.395pt}}
\multiput(198.58,138.00)(0.492,1.229){21}{\rule{0.119pt}{1.067pt}}
\multiput(197.17,138.00)(12.000,26.786){2}{\rule{0.400pt}{0.533pt}}
\multiput(210.58,167.00)(0.492,1.864){19}{\rule{0.118pt}{1.555pt}}
\multiput(209.17,167.00)(11.000,36.773){2}{\rule{0.400pt}{0.777pt}}
\multiput(221.58,207.00)(0.492,2.241){19}{\rule{0.118pt}{1.845pt}}
\multiput(220.17,207.00)(11.000,44.170){2}{\rule{0.400pt}{0.923pt}}
\multiput(232.58,255.00)(0.492,2.477){19}{\rule{0.118pt}{2.027pt}}
\multiput(231.17,255.00)(11.000,48.792){2}{\rule{0.400pt}{1.014pt}}
\multiput(243.58,308.00)(0.492,2.666){19}{\rule{0.118pt}{2.173pt}}
\multiput(242.17,308.00)(11.000,52.490){2}{\rule{0.400pt}{1.086pt}}
\multiput(254.58,365.00)(0.492,2.478){21}{\rule{0.119pt}{2.033pt}}
\multiput(253.17,365.00)(12.000,53.780){2}{\rule{0.400pt}{1.017pt}}
\multiput(266.58,423.00)(0.492,2.618){19}{\rule{0.118pt}{2.136pt}}
\multiput(265.17,423.00)(11.000,51.566){2}{\rule{0.400pt}{1.068pt}}
\multiput(277.58,479.00)(0.492,2.477){19}{\rule{0.118pt}{2.027pt}}
\multiput(276.17,479.00)(11.000,48.792){2}{\rule{0.400pt}{1.014pt}}
\multiput(288.58,532.00)(0.492,2.194){19}{\rule{0.118pt}{1.809pt}}
\multiput(287.17,532.00)(11.000,43.245){2}{\rule{0.400pt}{0.905pt}}
\multiput(299.58,579.00)(0.492,1.864){19}{\rule{0.118pt}{1.555pt}}
\multiput(298.17,579.00)(11.000,36.773){2}{\rule{0.400pt}{0.777pt}}
\multiput(310.58,619.00)(0.492,1.358){21}{\rule{0.119pt}{1.167pt}}
\multiput(309.17,619.00)(12.000,29.579){2}{\rule{0.400pt}{0.583pt}}
\multiput(322.58,651.00)(0.492,1.062){19}{\rule{0.118pt}{0.936pt}}
\multiput(321.17,651.00)(11.000,21.057){2}{\rule{0.400pt}{0.468pt}}
\multiput(333.58,674.00)(0.492,0.684){19}{\rule{0.118pt}{0.645pt}}
\multiput(332.17,674.00)(11.000,13.660){2}{\rule{0.400pt}{0.323pt}}
\multiput(344.00,689.59)(0.798,0.485){11}{\rule{0.729pt}{0.117pt}}
\multiput(344.00,688.17)(9.488,7.000){2}{\rule{0.364pt}{0.400pt}}
\multiput(366.00,694.93)(1.267,-0.477){7}{\rule{1.060pt}{0.115pt}}
\multiput(366.00,695.17)(9.800,-5.000){2}{\rule{0.530pt}{0.400pt}}
\multiput(378.00,689.92)(0.547,-0.491){17}{\rule{0.540pt}{0.118pt}}
\multiput(378.00,690.17)(9.879,-10.000){2}{\rule{0.270pt}{0.400pt}}
\multiput(389.00,679.92)(0.496,-0.492){19}{\rule{0.500pt}{0.118pt}}
\multiput(389.00,680.17)(9.962,-11.000){2}{\rule{0.250pt}{0.400pt}}
\multiput(400.58,667.77)(0.492,-0.543){19}{\rule{0.118pt}{0.536pt}}
\multiput(399.17,668.89)(11.000,-10.887){2}{\rule{0.400pt}{0.268pt}}
\multiput(411.00,656.92)(0.496,-0.492){19}{\rule{0.500pt}{0.118pt}}
\multiput(411.00,657.17)(9.962,-11.000){2}{\rule{0.250pt}{0.400pt}}
\multiput(422.00,645.93)(0.669,-0.489){15}{\rule{0.633pt}{0.118pt}}
\multiput(422.00,646.17)(10.685,-9.000){2}{\rule{0.317pt}{0.400pt}}
\multiput(434.00,636.93)(1.155,-0.477){7}{\rule{0.980pt}{0.115pt}}
\multiput(434.00,637.17)(8.966,-5.000){2}{\rule{0.490pt}{0.400pt}}
\put(445,631.67){\rule{2.650pt}{0.400pt}}
\multiput(445.00,632.17)(5.500,-1.000){2}{\rule{1.325pt}{0.400pt}}
\multiput(456.00,632.61)(2.248,0.447){3}{\rule{1.567pt}{0.108pt}}
\multiput(456.00,631.17)(7.748,3.000){2}{\rule{0.783pt}{0.400pt}}
\multiput(467.00,635.59)(0.798,0.485){11}{\rule{0.729pt}{0.117pt}}
\multiput(467.00,634.17)(9.488,7.000){2}{\rule{0.364pt}{0.400pt}}
\multiput(478.00,642.58)(0.600,0.491){17}{\rule{0.580pt}{0.118pt}}
\multiput(478.00,641.17)(10.796,10.000){2}{\rule{0.290pt}{0.400pt}}
\multiput(490.58,652.00)(0.492,0.543){19}{\rule{0.118pt}{0.536pt}}
\multiput(489.17,652.00)(11.000,10.887){2}{\rule{0.400pt}{0.268pt}}
\multiput(501.58,664.00)(0.492,0.543){19}{\rule{0.118pt}{0.536pt}}
\multiput(500.17,664.00)(11.000,10.887){2}{\rule{0.400pt}{0.268pt}}
\multiput(512.00,676.58)(0.547,0.491){17}{\rule{0.540pt}{0.118pt}}
\multiput(512.00,675.17)(9.879,10.000){2}{\rule{0.270pt}{0.400pt}}
\multiput(523.00,686.59)(0.692,0.488){13}{\rule{0.650pt}{0.117pt}}
\multiput(523.00,685.17)(9.651,8.000){2}{\rule{0.325pt}{0.400pt}}
\multiput(534.00,694.61)(2.472,0.447){3}{\rule{1.700pt}{0.108pt}}
\multiput(534.00,693.17)(8.472,3.000){2}{\rule{0.850pt}{0.400pt}}
\multiput(546.00,695.95)(2.248,-0.447){3}{\rule{1.567pt}{0.108pt}}
\multiput(546.00,696.17)(7.748,-3.000){2}{\rule{0.783pt}{0.400pt}}
\multiput(557.00,692.92)(0.496,-0.492){19}{\rule{0.500pt}{0.118pt}}
\multiput(557.00,693.17)(9.962,-11.000){2}{\rule{0.250pt}{0.400pt}}
\multiput(568.58,679.72)(0.492,-0.873){19}{\rule{0.118pt}{0.791pt}}
\multiput(567.17,681.36)(11.000,-17.358){2}{\rule{0.400pt}{0.395pt}}
\multiput(579.58,659.36)(0.492,-1.298){19}{\rule{0.118pt}{1.118pt}}
\multiput(578.17,661.68)(11.000,-25.679){2}{\rule{0.400pt}{0.559pt}}
\multiput(590.58,630.60)(0.492,-1.530){21}{\rule{0.119pt}{1.300pt}}
\multiput(589.17,633.30)(12.000,-33.302){2}{\rule{0.400pt}{0.650pt}}
\multiput(602.58,592.94)(0.492,-2.052){19}{\rule{0.118pt}{1.700pt}}
\multiput(601.17,596.47)(11.000,-40.472){2}{\rule{0.400pt}{0.850pt}}
\multiput(613.58,548.04)(0.492,-2.335){19}{\rule{0.118pt}{1.918pt}}
\multiput(612.17,552.02)(11.000,-46.019){2}{\rule{0.400pt}{0.959pt}}
\multiput(624.58,497.43)(0.492,-2.524){19}{\rule{0.118pt}{2.064pt}}
\multiput(623.17,501.72)(11.000,-49.717){2}{\rule{0.400pt}{1.032pt}}
\multiput(635.58,442.83)(0.492,-2.713){19}{\rule{0.118pt}{2.209pt}}
\multiput(634.17,447.41)(11.000,-53.415){2}{\rule{0.400pt}{1.105pt}}
\multiput(646.58,385.70)(0.492,-2.435){21}{\rule{0.119pt}{2.000pt}}
\multiput(645.17,389.85)(12.000,-52.849){2}{\rule{0.400pt}{1.000pt}}
\multiput(658.58,328.13)(0.492,-2.618){19}{\rule{0.118pt}{2.136pt}}
\multiput(657.17,332.57)(11.000,-51.566){2}{\rule{0.400pt}{1.068pt}}
\multiput(669.58,272.89)(0.492,-2.383){19}{\rule{0.118pt}{1.955pt}}
\multiput(668.17,276.94)(11.000,-46.943){2}{\rule{0.400pt}{0.977pt}}
\multiput(680.58,222.94)(0.492,-2.052){19}{\rule{0.118pt}{1.700pt}}
\multiput(679.17,226.47)(11.000,-40.472){2}{\rule{0.400pt}{0.850pt}}
\multiput(691.58,180.30)(0.492,-1.628){19}{\rule{0.118pt}{1.373pt}}
\multiput(690.17,183.15)(11.000,-32.151){2}{\rule{0.400pt}{0.686pt}}
\multiput(702.58,147.26)(0.492,-1.013){21}{\rule{0.119pt}{0.900pt}}
\multiput(701.17,149.13)(12.000,-22.132){2}{\rule{0.400pt}{0.450pt}}
\multiput(714.58,124.77)(0.492,-0.543){19}{\rule{0.118pt}{0.536pt}}
\multiput(713.17,125.89)(11.000,-10.887){2}{\rule{0.400pt}{0.268pt}}
\put(355.0,696.0){\rule[-0.200pt]{2.650pt}{0.400pt}}
\multiput(736.58,115.00)(0.492,0.543){19}{\rule{0.118pt}{0.536pt}}
\multiput(735.17,115.00)(11.000,10.887){2}{\rule{0.400pt}{0.268pt}}
\multiput(747.58,127.00)(0.492,1.013){21}{\rule{0.119pt}{0.900pt}}
\multiput(746.17,127.00)(12.000,22.132){2}{\rule{0.400pt}{0.450pt}}
\multiput(759.58,151.00)(0.492,1.628){19}{\rule{0.118pt}{1.373pt}}
\multiput(758.17,151.00)(11.000,32.151){2}{\rule{0.400pt}{0.686pt}}
\multiput(770.58,186.00)(0.492,2.052){19}{\rule{0.118pt}{1.700pt}}
\multiput(769.17,186.00)(11.000,40.472){2}{\rule{0.400pt}{0.850pt}}
\multiput(781.58,230.00)(0.492,2.383){19}{\rule{0.118pt}{1.955pt}}
\multiput(780.17,230.00)(11.000,46.943){2}{\rule{0.400pt}{0.977pt}}
\multiput(792.58,281.00)(0.492,2.618){19}{\rule{0.118pt}{2.136pt}}
\multiput(791.17,281.00)(11.000,51.566){2}{\rule{0.400pt}{1.068pt}}
\multiput(803.58,337.00)(0.492,2.435){21}{\rule{0.119pt}{2.000pt}}
\multiput(802.17,337.00)(12.000,52.849){2}{\rule{0.400pt}{1.000pt}}
\multiput(815.58,394.00)(0.492,2.713){19}{\rule{0.118pt}{2.209pt}}
\multiput(814.17,394.00)(11.000,53.415){2}{\rule{0.400pt}{1.105pt}}
\multiput(826.58,452.00)(0.492,2.524){19}{\rule{0.118pt}{2.064pt}}
\multiput(825.17,452.00)(11.000,49.717){2}{\rule{0.400pt}{1.032pt}}
\multiput(837.58,506.00)(0.492,2.335){19}{\rule{0.118pt}{1.918pt}}
\multiput(836.17,506.00)(11.000,46.019){2}{\rule{0.400pt}{0.959pt}}
\multiput(848.58,556.00)(0.492,2.052){19}{\rule{0.118pt}{1.700pt}}
\multiput(847.17,556.00)(11.000,40.472){2}{\rule{0.400pt}{0.850pt}}
\multiput(859.58,600.00)(0.492,1.530){21}{\rule{0.119pt}{1.300pt}}
\multiput(858.17,600.00)(12.000,33.302){2}{\rule{0.400pt}{0.650pt}}
\multiput(871.58,636.00)(0.492,1.298){19}{\rule{0.118pt}{1.118pt}}
\multiput(870.17,636.00)(11.000,25.679){2}{\rule{0.400pt}{0.559pt}}
\multiput(882.58,664.00)(0.492,0.873){19}{\rule{0.118pt}{0.791pt}}
\multiput(881.17,664.00)(11.000,17.358){2}{\rule{0.400pt}{0.395pt}}
\multiput(893.00,683.58)(0.496,0.492){19}{\rule{0.500pt}{0.118pt}}
\multiput(893.00,682.17)(9.962,11.000){2}{\rule{0.250pt}{0.400pt}}
\multiput(904.00,694.61)(2.248,0.447){3}{\rule{1.567pt}{0.108pt}}
\multiput(904.00,693.17)(7.748,3.000){2}{\rule{0.783pt}{0.400pt}}
\multiput(915.00,695.95)(2.472,-0.447){3}{\rule{1.700pt}{0.108pt}}
\multiput(915.00,696.17)(8.472,-3.000){2}{\rule{0.850pt}{0.400pt}}
\multiput(927.00,692.93)(0.692,-0.488){13}{\rule{0.650pt}{0.117pt}}
\multiput(927.00,693.17)(9.651,-8.000){2}{\rule{0.325pt}{0.400pt}}
\multiput(938.00,684.92)(0.547,-0.491){17}{\rule{0.540pt}{0.118pt}}
\multiput(938.00,685.17)(9.879,-10.000){2}{\rule{0.270pt}{0.400pt}}
\multiput(949.58,673.77)(0.492,-0.543){19}{\rule{0.118pt}{0.536pt}}
\multiput(948.17,674.89)(11.000,-10.887){2}{\rule{0.400pt}{0.268pt}}
\multiput(960.58,661.77)(0.492,-0.543){19}{\rule{0.118pt}{0.536pt}}
\multiput(959.17,662.89)(11.000,-10.887){2}{\rule{0.400pt}{0.268pt}}
\multiput(971.00,650.92)(0.600,-0.491){17}{\rule{0.580pt}{0.118pt}}
\multiput(971.00,651.17)(10.796,-10.000){2}{\rule{0.290pt}{0.400pt}}
\multiput(983.00,640.93)(0.798,-0.485){11}{\rule{0.729pt}{0.117pt}}
\multiput(983.00,641.17)(9.488,-7.000){2}{\rule{0.364pt}{0.400pt}}
\multiput(994.00,633.95)(2.248,-0.447){3}{\rule{1.567pt}{0.108pt}}
\multiput(994.00,634.17)(7.748,-3.000){2}{\rule{0.783pt}{0.400pt}}
\put(1005,631.67){\rule{2.650pt}{0.400pt}}
\multiput(1005.00,631.17)(5.500,1.000){2}{\rule{1.325pt}{0.400pt}}
\multiput(1016.00,633.59)(1.155,0.477){7}{\rule{0.980pt}{0.115pt}}
\multiput(1016.00,632.17)(8.966,5.000){2}{\rule{0.490pt}{0.400pt}}
\multiput(1027.00,638.59)(0.669,0.489){15}{\rule{0.633pt}{0.118pt}}
\multiput(1027.00,637.17)(10.685,9.000){2}{\rule{0.317pt}{0.400pt}}
\multiput(1039.00,647.58)(0.496,0.492){19}{\rule{0.500pt}{0.118pt}}
\multiput(1039.00,646.17)(9.962,11.000){2}{\rule{0.250pt}{0.400pt}}
\multiput(1050.58,658.00)(0.492,0.543){19}{\rule{0.118pt}{0.536pt}}
\multiput(1049.17,658.00)(11.000,10.887){2}{\rule{0.400pt}{0.268pt}}
\multiput(1061.00,670.58)(0.496,0.492){19}{\rule{0.500pt}{0.118pt}}
\multiput(1061.00,669.17)(9.962,11.000){2}{\rule{0.250pt}{0.400pt}}
\multiput(1072.00,681.58)(0.547,0.491){17}{\rule{0.540pt}{0.118pt}}
\multiput(1072.00,680.17)(9.879,10.000){2}{\rule{0.270pt}{0.400pt}}
\multiput(1083.00,691.59)(1.267,0.477){7}{\rule{1.060pt}{0.115pt}}
\multiput(1083.00,690.17)(9.800,5.000){2}{\rule{0.530pt}{0.400pt}}
\put(725.0,115.0){\rule[-0.200pt]{2.650pt}{0.400pt}}
\multiput(1106.00,694.93)(0.798,-0.485){11}{\rule{0.729pt}{0.117pt}}
\multiput(1106.00,695.17)(9.488,-7.000){2}{\rule{0.364pt}{0.400pt}}
\multiput(1117.58,686.32)(0.492,-0.684){19}{\rule{0.118pt}{0.645pt}}
\multiput(1116.17,687.66)(11.000,-13.660){2}{\rule{0.400pt}{0.323pt}}
\multiput(1128.58,670.11)(0.492,-1.062){19}{\rule{0.118pt}{0.936pt}}
\multiput(1127.17,672.06)(11.000,-21.057){2}{\rule{0.400pt}{0.468pt}}
\multiput(1139.58,646.16)(0.492,-1.358){21}{\rule{0.119pt}{1.167pt}}
\multiput(1138.17,648.58)(12.000,-29.579){2}{\rule{0.400pt}{0.583pt}}
\multiput(1151.58,612.55)(0.492,-1.864){19}{\rule{0.118pt}{1.555pt}}
\multiput(1150.17,615.77)(11.000,-36.773){2}{\rule{0.400pt}{0.777pt}}
\multiput(1162.58,571.49)(0.492,-2.194){19}{\rule{0.118pt}{1.809pt}}
\multiput(1161.17,575.25)(11.000,-43.245){2}{\rule{0.400pt}{0.905pt}}
\multiput(1173.58,523.58)(0.492,-2.477){19}{\rule{0.118pt}{2.027pt}}
\multiput(1172.17,527.79)(11.000,-48.792){2}{\rule{0.400pt}{1.014pt}}
\multiput(1184.58,470.13)(0.492,-2.618){19}{\rule{0.118pt}{2.136pt}}
\multiput(1183.17,474.57)(11.000,-51.566){2}{\rule{0.400pt}{1.068pt}}
\multiput(1195.58,414.56)(0.492,-2.478){21}{\rule{0.119pt}{2.033pt}}
\multiput(1194.17,418.78)(12.000,-53.780){2}{\rule{0.400pt}{1.017pt}}
\multiput(1207.58,355.98)(0.492,-2.666){19}{\rule{0.118pt}{2.173pt}}
\multiput(1206.17,360.49)(11.000,-52.490){2}{\rule{0.400pt}{1.086pt}}
\multiput(1218.58,299.58)(0.492,-2.477){19}{\rule{0.118pt}{2.027pt}}
\multiput(1217.17,303.79)(11.000,-48.792){2}{\rule{0.400pt}{1.014pt}}
\multiput(1229.58,247.34)(0.492,-2.241){19}{\rule{0.118pt}{1.845pt}}
\multiput(1228.17,251.17)(11.000,-44.170){2}{\rule{0.400pt}{0.923pt}}
\multiput(1240.58,200.55)(0.492,-1.864){19}{\rule{0.118pt}{1.555pt}}
\multiput(1239.17,203.77)(11.000,-36.773){2}{\rule{0.400pt}{0.777pt}}
\multiput(1251.58,162.57)(0.492,-1.229){21}{\rule{0.119pt}{1.067pt}}
\multiput(1250.17,164.79)(12.000,-26.786){2}{\rule{0.400pt}{0.533pt}}
\multiput(1263.58,134.72)(0.492,-0.873){19}{\rule{0.118pt}{0.791pt}}
\multiput(1262.17,136.36)(11.000,-17.358){2}{\rule{0.400pt}{0.395pt}}
\multiput(1274.00,117.93)(0.943,-0.482){9}{\rule{0.833pt}{0.116pt}}
\multiput(1274.00,118.17)(9.270,-6.000){2}{\rule{0.417pt}{0.400pt}}
\put(1095.0,696.0){\rule[-0.200pt]{2.650pt}{0.400pt}}
\sbox{\plotpoint}{\rule[-0.500pt]{1.000pt}{1.000pt}}%
\put(1059,218){\makebox(0,0)[r]{$B$}}
\multiput(1081,218)(20.756,0.000){4}{\usebox{\plotpoint}}
\put(1147,218){\usebox{\plotpoint}}
\put(176,632){\usebox{\plotpoint}}
\put(176.00,632.00){\usebox{\plotpoint}}
\put(196.76,632.00){\usebox{\plotpoint}}
\multiput(198,632)(20.756,0.000){0}{\usebox{\plotpoint}}
\put(217.51,632.00){\usebox{\plotpoint}}
\multiput(221,632)(20.756,0.000){0}{\usebox{\plotpoint}}
\put(238.27,632.00){\usebox{\plotpoint}}
\multiput(243,632)(20.756,0.000){0}{\usebox{\plotpoint}}
\put(259.02,632.00){\usebox{\plotpoint}}
\multiput(266,632)(20.756,0.000){0}{\usebox{\plotpoint}}
\put(279.78,632.00){\usebox{\plotpoint}}
\multiput(288,632)(20.756,0.000){0}{\usebox{\plotpoint}}
\put(300.53,632.00){\usebox{\plotpoint}}
\put(321.29,632.00){\usebox{\plotpoint}}
\multiput(322,632)(20.756,0.000){0}{\usebox{\plotpoint}}
\put(342.04,632.00){\usebox{\plotpoint}}
\multiput(344,632)(20.756,0.000){0}{\usebox{\plotpoint}}
\put(362.80,632.00){\usebox{\plotpoint}}
\multiput(366,632)(20.756,0.000){0}{\usebox{\plotpoint}}
\put(383.55,632.00){\usebox{\plotpoint}}
\multiput(389,632)(20.756,0.000){0}{\usebox{\plotpoint}}
\put(404.31,632.00){\usebox{\plotpoint}}
\multiput(411,632)(20.756,0.000){0}{\usebox{\plotpoint}}
\put(425.07,632.00){\usebox{\plotpoint}}
\multiput(434,632)(20.756,0.000){0}{\usebox{\plotpoint}}
\put(445.82,632.00){\usebox{\plotpoint}}
\put(466.58,632.00){\usebox{\plotpoint}}
\multiput(467,632)(20.756,0.000){0}{\usebox{\plotpoint}}
\put(487.33,632.00){\usebox{\plotpoint}}
\multiput(490,632)(20.756,0.000){0}{\usebox{\plotpoint}}
\put(508.09,632.00){\usebox{\plotpoint}}
\multiput(512,632)(20.756,0.000){0}{\usebox{\plotpoint}}
\put(528.84,632.00){\usebox{\plotpoint}}
\multiput(534,632)(20.756,0.000){0}{\usebox{\plotpoint}}
\put(549.60,632.00){\usebox{\plotpoint}}
\multiput(557,632)(20.756,0.000){0}{\usebox{\plotpoint}}
\put(570.35,632.00){\usebox{\plotpoint}}
\multiput(579,632)(20.756,0.000){0}{\usebox{\plotpoint}}
\put(591.11,632.00){\usebox{\plotpoint}}
\put(611.87,632.00){\usebox{\plotpoint}}
\multiput(613,632)(20.756,0.000){0}{\usebox{\plotpoint}}
\put(632.62,632.00){\usebox{\plotpoint}}
\multiput(635,632)(20.756,0.000){0}{\usebox{\plotpoint}}
\put(653.38,632.00){\usebox{\plotpoint}}
\multiput(658,632)(20.756,0.000){0}{\usebox{\plotpoint}}
\put(674.13,632.00){\usebox{\plotpoint}}
\multiput(680,632)(20.756,0.000){0}{\usebox{\plotpoint}}
\put(694.89,632.00){\usebox{\plotpoint}}
\multiput(702,632)(20.756,0.000){0}{\usebox{\plotpoint}}
\put(715.64,632.00){\usebox{\plotpoint}}
\multiput(725,632)(20.756,0.000){0}{\usebox{\plotpoint}}
\put(736.40,632.00){\usebox{\plotpoint}}
\put(757.15,632.00){\usebox{\plotpoint}}
\multiput(759,632)(20.756,0.000){0}{\usebox{\plotpoint}}
\put(777.91,632.00){\usebox{\plotpoint}}
\multiput(781,632)(20.756,0.000){0}{\usebox{\plotpoint}}
\put(798.66,632.00){\usebox{\plotpoint}}
\multiput(803,632)(20.756,0.000){0}{\usebox{\plotpoint}}
\put(819.42,632.00){\usebox{\plotpoint}}
\multiput(826,632)(20.756,0.000){0}{\usebox{\plotpoint}}
\put(840.18,632.00){\usebox{\plotpoint}}
\multiput(848,632)(20.756,0.000){0}{\usebox{\plotpoint}}
\put(860.93,632.00){\usebox{\plotpoint}}
\put(881.69,632.00){\usebox{\plotpoint}}
\multiput(882,632)(20.756,0.000){0}{\usebox{\plotpoint}}
\put(902.44,632.00){\usebox{\plotpoint}}
\multiput(904,632)(20.756,0.000){0}{\usebox{\plotpoint}}
\put(923.20,632.00){\usebox{\plotpoint}}
\multiput(927,632)(20.756,0.000){0}{\usebox{\plotpoint}}
\put(943.95,632.00){\usebox{\plotpoint}}
\multiput(949,632)(20.756,0.000){0}{\usebox{\plotpoint}}
\put(964.71,632.00){\usebox{\plotpoint}}
\multiput(971,632)(20.756,0.000){0}{\usebox{\plotpoint}}
\put(985.46,632.00){\usebox{\plotpoint}}
\multiput(994,632)(20.756,0.000){0}{\usebox{\plotpoint}}
\put(1006.22,632.00){\usebox{\plotpoint}}
\put(1026.98,632.00){\usebox{\plotpoint}}
\multiput(1027,632)(20.756,0.000){0}{\usebox{\plotpoint}}
\put(1047.73,632.00){\usebox{\plotpoint}}
\multiput(1050,632)(20.756,0.000){0}{\usebox{\plotpoint}}
\put(1068.49,632.00){\usebox{\plotpoint}}
\multiput(1072,632)(20.756,0.000){0}{\usebox{\plotpoint}}
\put(1089.24,632.00){\usebox{\plotpoint}}
\multiput(1095,632)(20.756,0.000){0}{\usebox{\plotpoint}}
\put(1110.00,632.00){\usebox{\plotpoint}}
\multiput(1117,632)(20.756,0.000){0}{\usebox{\plotpoint}}
\put(1130.75,632.00){\usebox{\plotpoint}}
\multiput(1139,632)(20.756,0.000){0}{\usebox{\plotpoint}}
\put(1151.51,632.00){\usebox{\plotpoint}}
\put(1172.26,632.00){\usebox{\plotpoint}}
\multiput(1173,632)(20.756,0.000){0}{\usebox{\plotpoint}}
\put(1193.02,632.00){\usebox{\plotpoint}}
\multiput(1195,632)(20.756,0.000){0}{\usebox{\plotpoint}}
\put(1213.77,632.00){\usebox{\plotpoint}}
\multiput(1218,632)(20.756,0.000){0}{\usebox{\plotpoint}}
\put(1234.53,632.00){\usebox{\plotpoint}}
\multiput(1240,632)(20.756,0.000){0}{\usebox{\plotpoint}}
\put(1255.29,632.00){\usebox{\plotpoint}}
\multiput(1263,632)(20.756,0.000){0}{\usebox{\plotpoint}}
\put(1276.04,632.00){\usebox{\plotpoint}}
\put(1285,632){\usebox{\plotpoint}}
\end{picture}

%% file: fig2.tex
% GNUPLOT: LaTeX picture
\setlength{\unitlength}{0.240900pt}
\ifx\plotpoint\undefined\newsavebox{\plotpoint}\fi
\sbox{\plotpoint}{\rule[-0.200pt]{0.400pt}{0.400pt}}%
\begin{picture}(1349,720)(0,0)
\font\gnuplot=cmr10 at 10pt
\gnuplot
\sbox{\plotpoint}{\rule[-0.200pt]{0.400pt}{0.400pt}}%
\put(220.0,113.0){\rule[-0.200pt]{256.558pt}{0.400pt}}
\put(220.0,113.0){\rule[-0.200pt]{0.400pt}{140.686pt}}
\put(220.0,113.0){\rule[-0.200pt]{4.818pt}{0.400pt}}
\put(198,113){\makebox(0,0)[r]{0}}
\put(1265.0,113.0){\rule[-0.200pt]{4.818pt}{0.400pt}}
\put(220.0,219.0){\rule[-0.200pt]{4.818pt}{0.400pt}}
\put(198,219){\makebox(0,0)[r]{0.1}}
\put(1265.0,219.0){\rule[-0.200pt]{4.818pt}{0.400pt}}
\put(220.0,325.0){\rule[-0.200pt]{4.818pt}{0.400pt}}
\put(198,325){\makebox(0,0)[r]{0.2}}
\put(1265.0,325.0){\rule[-0.200pt]{4.818pt}{0.400pt}}
\put(220.0,432.0){\rule[-0.200pt]{4.818pt}{0.400pt}}
\put(198,432){\makebox(0,0)[r]{0.3}}
\put(1265.0,432.0){\rule[-0.200pt]{4.818pt}{0.400pt}}
\put(220.0,538.0){\rule[-0.200pt]{4.818pt}{0.400pt}}
\put(198,538){\makebox(0,0)[r]{0.4}}
\put(1265.0,538.0){\rule[-0.200pt]{4.818pt}{0.400pt}}
\put(220.0,644.0){\rule[-0.200pt]{4.818pt}{0.400pt}}
\put(198,644){\makebox(0,0)[r]{0.5}}
\put(1265.0,644.0){\rule[-0.200pt]{4.818pt}{0.400pt}}
\put(220.0,113.0){\rule[-0.200pt]{0.400pt}{4.818pt}}
\put(220,68){\makebox(0,0){0}}
\put(220.0,677.0){\rule[-0.200pt]{0.400pt}{4.818pt}}
\put(433.0,113.0){\rule[-0.200pt]{0.400pt}{4.818pt}}
\put(433,68){\makebox(0,0){0.2}}
\put(433.0,677.0){\rule[-0.200pt]{0.400pt}{4.818pt}}
\put(646.0,113.0){\rule[-0.200pt]{0.400pt}{4.818pt}}
\put(646,68){\makebox(0,0){0.4}}
\put(646.0,677.0){\rule[-0.200pt]{0.400pt}{4.818pt}}
\put(859.0,113.0){\rule[-0.200pt]{0.400pt}{4.818pt}}
\put(859,68){\makebox(0,0){0.6}}
\put(859.0,677.0){\rule[-0.200pt]{0.400pt}{4.818pt}}
\put(1072.0,113.0){\rule[-0.200pt]{0.400pt}{4.818pt}}
\put(1072,68){\makebox(0,0){0.8}}
\put(1072.0,677.0){\rule[-0.200pt]{0.400pt}{4.818pt}}
\put(1285.0,113.0){\rule[-0.200pt]{0.400pt}{4.818pt}}
\put(1285,68){\makebox(0,0){1}}
\put(1285.0,677.0){\rule[-0.200pt]{0.400pt}{4.818pt}}
\put(220.0,113.0){\rule[-0.200pt]{256.558pt}{0.400pt}}
\put(1285.0,113.0){\rule[-0.200pt]{0.400pt}{140.686pt}}
\put(220.0,697.0){\rule[-0.200pt]{256.558pt}{0.400pt}}
\put(45,405){\makebox(0,0){$\Delta B_{\rm max}$}}
\put(752,-22){\makebox(0,0){$\varepsilon$}}
\put(220.0,113.0){\rule[-0.200pt]{0.400pt}{140.686pt}}
\put(1155,632){\makebox(0,0)[r]{Paz, Mahler [4]}}
\put(1177.0,632.0){\rule[-0.200pt]{15.899pt}{0.400pt}}
\put(220,553){\usebox{\plotpoint}}
\put(220,551.67){\rule{2.650pt}{0.400pt}}
\multiput(220.00,552.17)(5.500,-1.000){2}{\rule{1.325pt}{0.400pt}}
\put(231,550.67){\rule{2.650pt}{0.400pt}}
\multiput(231.00,551.17)(5.500,-1.000){2}{\rule{1.325pt}{0.400pt}}
\put(242,549.17){\rule{2.100pt}{0.400pt}}
\multiput(242.00,550.17)(5.641,-2.000){2}{\rule{1.050pt}{0.400pt}}
\put(252,547.67){\rule{2.650pt}{0.400pt}}
\multiput(252.00,548.17)(5.500,-1.000){2}{\rule{1.325pt}{0.400pt}}
\multiput(263.00,546.95)(2.248,-0.447){3}{\rule{1.567pt}{0.108pt}}
\multiput(263.00,547.17)(7.748,-3.000){2}{\rule{0.783pt}{0.400pt}}
\put(274,543.17){\rule{2.300pt}{0.400pt}}
\multiput(274.00,544.17)(6.226,-2.000){2}{\rule{1.150pt}{0.400pt}}
\put(285,541.17){\rule{2.100pt}{0.400pt}}
\multiput(285.00,542.17)(5.641,-2.000){2}{\rule{1.050pt}{0.400pt}}
\multiput(295.00,539.95)(2.248,-0.447){3}{\rule{1.567pt}{0.108pt}}
\multiput(295.00,540.17)(7.748,-3.000){2}{\rule{0.783pt}{0.400pt}}
\multiput(306.00,536.95)(2.248,-0.447){3}{\rule{1.567pt}{0.108pt}}
\multiput(306.00,537.17)(7.748,-3.000){2}{\rule{0.783pt}{0.400pt}}
\multiput(317.00,533.95)(2.248,-0.447){3}{\rule{1.567pt}{0.108pt}}
\multiput(317.00,534.17)(7.748,-3.000){2}{\rule{0.783pt}{0.400pt}}
\multiput(328.00,530.94)(1.358,-0.468){5}{\rule{1.100pt}{0.113pt}}
\multiput(328.00,531.17)(7.717,-4.000){2}{\rule{0.550pt}{0.400pt}}
\multiput(338.00,526.95)(2.248,-0.447){3}{\rule{1.567pt}{0.108pt}}
\multiput(338.00,527.17)(7.748,-3.000){2}{\rule{0.783pt}{0.400pt}}
\multiput(349.00,523.94)(1.505,-0.468){5}{\rule{1.200pt}{0.113pt}}
\multiput(349.00,524.17)(8.509,-4.000){2}{\rule{0.600pt}{0.400pt}}
\multiput(360.00,519.94)(1.505,-0.468){5}{\rule{1.200pt}{0.113pt}}
\multiput(360.00,520.17)(8.509,-4.000){2}{\rule{0.600pt}{0.400pt}}
\multiput(371.00,515.94)(1.358,-0.468){5}{\rule{1.100pt}{0.113pt}}
\multiput(371.00,516.17)(7.717,-4.000){2}{\rule{0.550pt}{0.400pt}}
\multiput(381.00,511.94)(1.505,-0.468){5}{\rule{1.200pt}{0.113pt}}
\multiput(381.00,512.17)(8.509,-4.000){2}{\rule{0.600pt}{0.400pt}}
\multiput(392.00,507.94)(1.505,-0.468){5}{\rule{1.200pt}{0.113pt}}
\multiput(392.00,508.17)(8.509,-4.000){2}{\rule{0.600pt}{0.400pt}}
\multiput(403.00,503.93)(1.155,-0.477){7}{\rule{0.980pt}{0.115pt}}
\multiput(403.00,504.17)(8.966,-5.000){2}{\rule{0.490pt}{0.400pt}}
\multiput(414.00,498.94)(1.358,-0.468){5}{\rule{1.100pt}{0.113pt}}
\multiput(414.00,499.17)(7.717,-4.000){2}{\rule{0.550pt}{0.400pt}}
\multiput(424.00,494.93)(1.155,-0.477){7}{\rule{0.980pt}{0.115pt}}
\multiput(424.00,495.17)(8.966,-5.000){2}{\rule{0.490pt}{0.400pt}}
\multiput(435.00,489.93)(1.155,-0.477){7}{\rule{0.980pt}{0.115pt}}
\multiput(435.00,490.17)(8.966,-5.000){2}{\rule{0.490pt}{0.400pt}}
\multiput(446.00,484.93)(1.155,-0.477){7}{\rule{0.980pt}{0.115pt}}
\multiput(446.00,485.17)(8.966,-5.000){2}{\rule{0.490pt}{0.400pt}}
\multiput(457.00,479.93)(1.044,-0.477){7}{\rule{0.900pt}{0.115pt}}
\multiput(457.00,480.17)(8.132,-5.000){2}{\rule{0.450pt}{0.400pt}}
\multiput(467.00,474.93)(0.943,-0.482){9}{\rule{0.833pt}{0.116pt}}
\multiput(467.00,475.17)(9.270,-6.000){2}{\rule{0.417pt}{0.400pt}}
\multiput(478.00,468.93)(1.155,-0.477){7}{\rule{0.980pt}{0.115pt}}
\multiput(478.00,469.17)(8.966,-5.000){2}{\rule{0.490pt}{0.400pt}}
\multiput(489.00,463.93)(0.943,-0.482){9}{\rule{0.833pt}{0.116pt}}
\multiput(489.00,464.17)(9.270,-6.000){2}{\rule{0.417pt}{0.400pt}}
\multiput(500.00,457.93)(0.852,-0.482){9}{\rule{0.767pt}{0.116pt}}
\multiput(500.00,458.17)(8.409,-6.000){2}{\rule{0.383pt}{0.400pt}}
\multiput(510.00,451.93)(0.943,-0.482){9}{\rule{0.833pt}{0.116pt}}
\multiput(510.00,452.17)(9.270,-6.000){2}{\rule{0.417pt}{0.400pt}}
\multiput(521.00,445.93)(0.943,-0.482){9}{\rule{0.833pt}{0.116pt}}
\multiput(521.00,446.17)(9.270,-6.000){2}{\rule{0.417pt}{0.400pt}}
\multiput(532.00,439.93)(0.943,-0.482){9}{\rule{0.833pt}{0.116pt}}
\multiput(532.00,440.17)(9.270,-6.000){2}{\rule{0.417pt}{0.400pt}}
\multiput(543.00,433.93)(0.721,-0.485){11}{\rule{0.671pt}{0.117pt}}
\multiput(543.00,434.17)(8.606,-7.000){2}{\rule{0.336pt}{0.400pt}}
\multiput(553.00,426.93)(0.943,-0.482){9}{\rule{0.833pt}{0.116pt}}
\multiput(553.00,427.17)(9.270,-6.000){2}{\rule{0.417pt}{0.400pt}}
\multiput(564.00,420.93)(0.798,-0.485){11}{\rule{0.729pt}{0.117pt}}
\multiput(564.00,421.17)(9.488,-7.000){2}{\rule{0.364pt}{0.400pt}}
\multiput(575.00,413.93)(0.798,-0.485){11}{\rule{0.729pt}{0.117pt}}
\multiput(575.00,414.17)(9.488,-7.000){2}{\rule{0.364pt}{0.400pt}}
\multiput(586.00,406.93)(0.798,-0.485){11}{\rule{0.729pt}{0.117pt}}
\multiput(586.00,407.17)(9.488,-7.000){2}{\rule{0.364pt}{0.400pt}}
\multiput(597.00,399.93)(0.721,-0.485){11}{\rule{0.671pt}{0.117pt}}
\multiput(597.00,400.17)(8.606,-7.000){2}{\rule{0.336pt}{0.400pt}}
\multiput(607.00,392.93)(0.798,-0.485){11}{\rule{0.729pt}{0.117pt}}
\multiput(607.00,393.17)(9.488,-7.000){2}{\rule{0.364pt}{0.400pt}}
\multiput(618.00,385.93)(0.692,-0.488){13}{\rule{0.650pt}{0.117pt}}
\multiput(618.00,386.17)(9.651,-8.000){2}{\rule{0.325pt}{0.400pt}}
\multiput(629.00,377.93)(0.692,-0.488){13}{\rule{0.650pt}{0.117pt}}
\multiput(629.00,378.17)(9.651,-8.000){2}{\rule{0.325pt}{0.400pt}}
\multiput(640.00,369.93)(0.626,-0.488){13}{\rule{0.600pt}{0.117pt}}
\multiput(640.00,370.17)(8.755,-8.000){2}{\rule{0.300pt}{0.400pt}}
\multiput(650.00,361.93)(0.692,-0.488){13}{\rule{0.650pt}{0.117pt}}
\multiput(650.00,362.17)(9.651,-8.000){2}{\rule{0.325pt}{0.400pt}}
\multiput(661.00,353.93)(0.692,-0.488){13}{\rule{0.650pt}{0.117pt}}
\multiput(661.00,354.17)(9.651,-8.000){2}{\rule{0.325pt}{0.400pt}}
\multiput(672.00,345.93)(0.692,-0.488){13}{\rule{0.650pt}{0.117pt}}
\multiput(672.00,346.17)(9.651,-8.000){2}{\rule{0.325pt}{0.400pt}}
\multiput(683.00,337.93)(0.553,-0.489){15}{\rule{0.544pt}{0.118pt}}
\multiput(683.00,338.17)(8.870,-9.000){2}{\rule{0.272pt}{0.400pt}}
\multiput(693.00,328.93)(0.692,-0.488){13}{\rule{0.650pt}{0.117pt}}
\multiput(693.00,329.17)(9.651,-8.000){2}{\rule{0.325pt}{0.400pt}}
\multiput(704.00,320.93)(0.611,-0.489){15}{\rule{0.589pt}{0.118pt}}
\multiput(704.00,321.17)(9.778,-9.000){2}{\rule{0.294pt}{0.400pt}}
\multiput(715.00,311.93)(0.611,-0.489){15}{\rule{0.589pt}{0.118pt}}
\multiput(715.00,312.17)(9.778,-9.000){2}{\rule{0.294pt}{0.400pt}}
\multiput(726.00,302.92)(0.495,-0.491){17}{\rule{0.500pt}{0.118pt}}
\multiput(726.00,303.17)(8.962,-10.000){2}{\rule{0.250pt}{0.400pt}}
\multiput(736.00,292.93)(0.611,-0.489){15}{\rule{0.589pt}{0.118pt}}
\multiput(736.00,293.17)(9.778,-9.000){2}{\rule{0.294pt}{0.400pt}}
\multiput(747.00,283.92)(0.547,-0.491){17}{\rule{0.540pt}{0.118pt}}
\multiput(747.00,284.17)(9.879,-10.000){2}{\rule{0.270pt}{0.400pt}}
\multiput(758.00,273.92)(0.547,-0.491){17}{\rule{0.540pt}{0.118pt}}
\multiput(758.00,274.17)(9.879,-10.000){2}{\rule{0.270pt}{0.400pt}}
\multiput(769.00,263.92)(0.495,-0.491){17}{\rule{0.500pt}{0.118pt}}
\multiput(769.00,264.17)(8.962,-10.000){2}{\rule{0.250pt}{0.400pt}}
\multiput(779.00,253.92)(0.547,-0.491){17}{\rule{0.540pt}{0.118pt}}
\multiput(779.00,254.17)(9.879,-10.000){2}{\rule{0.270pt}{0.400pt}}
\multiput(790.00,243.92)(0.496,-0.492){19}{\rule{0.500pt}{0.118pt}}
\multiput(790.00,244.17)(9.962,-11.000){2}{\rule{0.250pt}{0.400pt}}
\multiput(801.00,232.92)(0.547,-0.491){17}{\rule{0.540pt}{0.118pt}}
\multiput(801.00,233.17)(9.879,-10.000){2}{\rule{0.270pt}{0.400pt}}
\multiput(812.58,221.76)(0.491,-0.547){17}{\rule{0.118pt}{0.540pt}}
\multiput(811.17,222.88)(10.000,-9.879){2}{\rule{0.400pt}{0.270pt}}
\multiput(822.00,211.92)(0.496,-0.492){19}{\rule{0.500pt}{0.118pt}}
\multiput(822.00,212.17)(9.962,-11.000){2}{\rule{0.250pt}{0.400pt}}
\multiput(833.58,199.77)(0.492,-0.543){19}{\rule{0.118pt}{0.536pt}}
\multiput(832.17,200.89)(11.000,-10.887){2}{\rule{0.400pt}{0.268pt}}
\multiput(844.00,188.92)(0.496,-0.492){19}{\rule{0.500pt}{0.118pt}}
\multiput(844.00,189.17)(9.962,-11.000){2}{\rule{0.250pt}{0.400pt}}
\multiput(855.58,176.59)(0.491,-0.600){17}{\rule{0.118pt}{0.580pt}}
\multiput(854.17,177.80)(10.000,-10.796){2}{\rule{0.400pt}{0.290pt}}
\multiput(865.58,164.77)(0.492,-0.543){19}{\rule{0.118pt}{0.536pt}}
\multiput(864.17,165.89)(11.000,-10.887){2}{\rule{0.400pt}{0.268pt}}
\multiput(876.58,152.62)(0.492,-0.590){19}{\rule{0.118pt}{0.573pt}}
\multiput(875.17,153.81)(11.000,-11.811){2}{\rule{0.400pt}{0.286pt}}
\multiput(887.58,139.77)(0.492,-0.543){19}{\rule{0.118pt}{0.536pt}}
\multiput(886.17,140.89)(11.000,-10.887){2}{\rule{0.400pt}{0.268pt}}
\multiput(898.58,127.43)(0.491,-0.652){17}{\rule{0.118pt}{0.620pt}}
\multiput(897.17,128.71)(10.000,-11.713){2}{\rule{0.400pt}{0.310pt}}
\multiput(908.00,115.94)(0.481,-0.468){5}{\rule{0.500pt}{0.113pt}}
\multiput(908.00,116.17)(2.962,-4.000){2}{\rule{0.250pt}{0.400pt}}
\sbox{\plotpoint}{\rule[-0.400pt]{0.800pt}{0.800pt}}%
\put(1155,587){\makebox(0,0)[r]{Huelga, Marshall, Santos [5]}}
\put(1177.0,587.0){\rule[-0.400pt]{15.899pt}{0.800pt}}
\put(220,644){\usebox{\plotpoint}}
\put(220,641.84){\rule{2.650pt}{0.800pt}}
\multiput(220.00,642.34)(5.500,-1.000){2}{\rule{1.325pt}{0.800pt}}
\put(231,640.84){\rule{2.650pt}{0.800pt}}
\multiput(231.00,641.34)(5.500,-1.000){2}{\rule{1.325pt}{0.800pt}}
\put(242,639.34){\rule{2.409pt}{0.800pt}}
\multiput(242.00,640.34)(5.000,-2.000){2}{\rule{1.204pt}{0.800pt}}
\put(252,637.34){\rule{2.650pt}{0.800pt}}
\multiput(252.00,638.34)(5.500,-2.000){2}{\rule{1.325pt}{0.800pt}}
\put(263,635.34){\rule{2.650pt}{0.800pt}}
\multiput(263.00,636.34)(5.500,-2.000){2}{\rule{1.325pt}{0.800pt}}
\put(274,633.34){\rule{2.650pt}{0.800pt}}
\multiput(274.00,634.34)(5.500,-2.000){2}{\rule{1.325pt}{0.800pt}}
\put(285,630.84){\rule{2.409pt}{0.800pt}}
\multiput(285.00,632.34)(5.000,-3.000){2}{\rule{1.204pt}{0.800pt}}
\put(295,627.84){\rule{2.650pt}{0.800pt}}
\multiput(295.00,629.34)(5.500,-3.000){2}{\rule{1.325pt}{0.800pt}}
\put(306,624.84){\rule{2.650pt}{0.800pt}}
\multiput(306.00,626.34)(5.500,-3.000){2}{\rule{1.325pt}{0.800pt}}
\put(317,621.84){\rule{2.650pt}{0.800pt}}
\multiput(317.00,623.34)(5.500,-3.000){2}{\rule{1.325pt}{0.800pt}}
\put(328,618.34){\rule{2.200pt}{0.800pt}}
\multiput(328.00,620.34)(5.434,-4.000){2}{\rule{1.100pt}{0.800pt}}
\put(338,614.34){\rule{2.400pt}{0.800pt}}
\multiput(338.00,616.34)(6.019,-4.000){2}{\rule{1.200pt}{0.800pt}}
\put(349,610.34){\rule{2.400pt}{0.800pt}}
\multiput(349.00,612.34)(6.019,-4.000){2}{\rule{1.200pt}{0.800pt}}
\put(360,606.34){\rule{2.400pt}{0.800pt}}
\multiput(360.00,608.34)(6.019,-4.000){2}{\rule{1.200pt}{0.800pt}}
\put(371,602.34){\rule{2.200pt}{0.800pt}}
\multiput(371.00,604.34)(5.434,-4.000){2}{\rule{1.100pt}{0.800pt}}
\put(381,598.34){\rule{2.400pt}{0.800pt}}
\multiput(381.00,600.34)(6.019,-4.000){2}{\rule{1.200pt}{0.800pt}}
\multiput(392.00,596.06)(1.432,-0.560){3}{\rule{1.960pt}{0.135pt}}
\multiput(392.00,596.34)(6.932,-5.000){2}{\rule{0.980pt}{0.800pt}}
\multiput(403.00,591.06)(1.432,-0.560){3}{\rule{1.960pt}{0.135pt}}
\multiput(403.00,591.34)(6.932,-5.000){2}{\rule{0.980pt}{0.800pt}}
\multiput(414.00,586.06)(1.264,-0.560){3}{\rule{1.800pt}{0.135pt}}
\multiput(414.00,586.34)(6.264,-5.000){2}{\rule{0.900pt}{0.800pt}}
\multiput(424.00,581.06)(1.432,-0.560){3}{\rule{1.960pt}{0.135pt}}
\multiput(424.00,581.34)(6.932,-5.000){2}{\rule{0.980pt}{0.800pt}}
\multiput(435.00,576.06)(1.432,-0.560){3}{\rule{1.960pt}{0.135pt}}
\multiput(435.00,576.34)(6.932,-5.000){2}{\rule{0.980pt}{0.800pt}}
\multiput(446.00,571.06)(1.432,-0.560){3}{\rule{1.960pt}{0.135pt}}
\multiput(446.00,571.34)(6.932,-5.000){2}{\rule{0.980pt}{0.800pt}}
\multiput(457.00,566.07)(0.909,-0.536){5}{\rule{1.533pt}{0.129pt}}
\multiput(457.00,566.34)(6.817,-6.000){2}{\rule{0.767pt}{0.800pt}}
\multiput(467.00,560.07)(1.020,-0.536){5}{\rule{1.667pt}{0.129pt}}
\multiput(467.00,560.34)(7.541,-6.000){2}{\rule{0.833pt}{0.800pt}}
\multiput(478.00,554.06)(1.432,-0.560){3}{\rule{1.960pt}{0.135pt}}
\multiput(478.00,554.34)(6.932,-5.000){2}{\rule{0.980pt}{0.800pt}}
\multiput(489.00,549.07)(1.020,-0.536){5}{\rule{1.667pt}{0.129pt}}
\multiput(489.00,549.34)(7.541,-6.000){2}{\rule{0.833pt}{0.800pt}}
\multiput(500.00,543.08)(0.738,-0.526){7}{\rule{1.343pt}{0.127pt}}
\multiput(500.00,543.34)(7.213,-7.000){2}{\rule{0.671pt}{0.800pt}}
\multiput(510.00,536.07)(1.020,-0.536){5}{\rule{1.667pt}{0.129pt}}
\multiput(510.00,536.34)(7.541,-6.000){2}{\rule{0.833pt}{0.800pt}}
\multiput(521.00,530.08)(0.825,-0.526){7}{\rule{1.457pt}{0.127pt}}
\multiput(521.00,530.34)(7.976,-7.000){2}{\rule{0.729pt}{0.800pt}}
\multiput(532.00,523.07)(1.020,-0.536){5}{\rule{1.667pt}{0.129pt}}
\multiput(532.00,523.34)(7.541,-6.000){2}{\rule{0.833pt}{0.800pt}}
\multiput(543.00,517.08)(0.738,-0.526){7}{\rule{1.343pt}{0.127pt}}
\multiput(543.00,517.34)(7.213,-7.000){2}{\rule{0.671pt}{0.800pt}}
\multiput(553.00,510.08)(0.825,-0.526){7}{\rule{1.457pt}{0.127pt}}
\multiput(553.00,510.34)(7.976,-7.000){2}{\rule{0.729pt}{0.800pt}}
\multiput(564.00,503.08)(0.825,-0.526){7}{\rule{1.457pt}{0.127pt}}
\multiput(564.00,503.34)(7.976,-7.000){2}{\rule{0.729pt}{0.800pt}}
\multiput(575.00,496.08)(0.700,-0.520){9}{\rule{1.300pt}{0.125pt}}
\multiput(575.00,496.34)(8.302,-8.000){2}{\rule{0.650pt}{0.800pt}}
\multiput(586.00,488.08)(0.825,-0.526){7}{\rule{1.457pt}{0.127pt}}
\multiput(586.00,488.34)(7.976,-7.000){2}{\rule{0.729pt}{0.800pt}}
\multiput(597.00,481.08)(0.627,-0.520){9}{\rule{1.200pt}{0.125pt}}
\multiput(597.00,481.34)(7.509,-8.000){2}{\rule{0.600pt}{0.800pt}}
\multiput(607.00,473.08)(0.825,-0.526){7}{\rule{1.457pt}{0.127pt}}
\multiput(607.00,473.34)(7.976,-7.000){2}{\rule{0.729pt}{0.800pt}}
\multiput(618.00,466.08)(0.700,-0.520){9}{\rule{1.300pt}{0.125pt}}
\multiput(618.00,466.34)(8.302,-8.000){2}{\rule{0.650pt}{0.800pt}}
\multiput(629.00,458.08)(0.611,-0.516){11}{\rule{1.178pt}{0.124pt}}
\multiput(629.00,458.34)(8.555,-9.000){2}{\rule{0.589pt}{0.800pt}}
\multiput(640.00,449.08)(0.627,-0.520){9}{\rule{1.200pt}{0.125pt}}
\multiput(640.00,449.34)(7.509,-8.000){2}{\rule{0.600pt}{0.800pt}}
\multiput(650.00,441.08)(0.611,-0.516){11}{\rule{1.178pt}{0.124pt}}
\multiput(650.00,441.34)(8.555,-9.000){2}{\rule{0.589pt}{0.800pt}}
\multiput(661.00,432.08)(0.700,-0.520){9}{\rule{1.300pt}{0.125pt}}
\multiput(661.00,432.34)(8.302,-8.000){2}{\rule{0.650pt}{0.800pt}}
\multiput(672.00,424.08)(0.611,-0.516){11}{\rule{1.178pt}{0.124pt}}
\multiput(672.00,424.34)(8.555,-9.000){2}{\rule{0.589pt}{0.800pt}}
\multiput(683.00,415.08)(0.548,-0.516){11}{\rule{1.089pt}{0.124pt}}
\multiput(683.00,415.34)(7.740,-9.000){2}{\rule{0.544pt}{0.800pt}}
\multiput(693.00,406.08)(0.611,-0.516){11}{\rule{1.178pt}{0.124pt}}
\multiput(693.00,406.34)(8.555,-9.000){2}{\rule{0.589pt}{0.800pt}}
\multiput(704.00,397.08)(0.543,-0.514){13}{\rule{1.080pt}{0.124pt}}
\multiput(704.00,397.34)(8.758,-10.000){2}{\rule{0.540pt}{0.800pt}}
\multiput(715.00,387.08)(0.543,-0.514){13}{\rule{1.080pt}{0.124pt}}
\multiput(715.00,387.34)(8.758,-10.000){2}{\rule{0.540pt}{0.800pt}}
\multiput(726.00,377.08)(0.548,-0.516){11}{\rule{1.089pt}{0.124pt}}
\multiput(726.00,377.34)(7.740,-9.000){2}{\rule{0.544pt}{0.800pt}}
\multiput(736.00,368.08)(0.543,-0.514){13}{\rule{1.080pt}{0.124pt}}
\multiput(736.00,368.34)(8.758,-10.000){2}{\rule{0.540pt}{0.800pt}}
\multiput(747.00,358.08)(0.489,-0.512){15}{\rule{1.000pt}{0.123pt}}
\multiput(747.00,358.34)(8.924,-11.000){2}{\rule{0.500pt}{0.800pt}}
\multiput(758.00,347.08)(0.543,-0.514){13}{\rule{1.080pt}{0.124pt}}
\multiput(758.00,347.34)(8.758,-10.000){2}{\rule{0.540pt}{0.800pt}}
\multiput(770.40,334.52)(0.514,-0.543){13}{\rule{0.124pt}{1.080pt}}
\multiput(767.34,336.76)(10.000,-8.758){2}{\rule{0.800pt}{0.540pt}}
\multiput(779.00,326.08)(0.489,-0.512){15}{\rule{1.000pt}{0.123pt}}
\multiput(779.00,326.34)(8.924,-11.000){2}{\rule{0.500pt}{0.800pt}}
\multiput(790.00,315.08)(0.489,-0.512){15}{\rule{1.000pt}{0.123pt}}
\multiput(790.00,315.34)(8.924,-11.000){2}{\rule{0.500pt}{0.800pt}}
\multiput(801.00,304.08)(0.489,-0.512){15}{\rule{1.000pt}{0.123pt}}
\multiput(801.00,304.34)(8.924,-11.000){2}{\rule{0.500pt}{0.800pt}}
\multiput(813.40,290.18)(0.514,-0.599){13}{\rule{0.124pt}{1.160pt}}
\multiput(810.34,292.59)(10.000,-9.592){2}{\rule{0.800pt}{0.580pt}}
\multiput(822.00,281.08)(0.489,-0.512){15}{\rule{1.000pt}{0.123pt}}
\multiput(822.00,281.34)(8.924,-11.000){2}{\rule{0.500pt}{0.800pt}}
\multiput(834.40,267.55)(0.512,-0.539){15}{\rule{0.123pt}{1.073pt}}
\multiput(831.34,269.77)(11.000,-9.774){2}{\rule{0.800pt}{0.536pt}}
\multiput(845.40,255.25)(0.512,-0.589){15}{\rule{0.123pt}{1.145pt}}
\multiput(842.34,257.62)(11.000,-10.623){2}{\rule{0.800pt}{0.573pt}}
\multiput(856.40,242.18)(0.514,-0.599){13}{\rule{0.124pt}{1.160pt}}
\multiput(853.34,244.59)(10.000,-9.592){2}{\rule{0.800pt}{0.580pt}}
\multiput(866.40,230.25)(0.512,-0.589){15}{\rule{0.123pt}{1.145pt}}
\multiput(863.34,232.62)(11.000,-10.623){2}{\rule{0.800pt}{0.573pt}}
\multiput(877.40,217.25)(0.512,-0.589){15}{\rule{0.123pt}{1.145pt}}
\multiput(874.34,219.62)(11.000,-10.623){2}{\rule{0.800pt}{0.573pt}}
\multiput(888.40,203.94)(0.512,-0.639){15}{\rule{0.123pt}{1.218pt}}
\multiput(885.34,206.47)(11.000,-11.472){2}{\rule{0.800pt}{0.609pt}}
\multiput(899.40,189.85)(0.514,-0.654){13}{\rule{0.124pt}{1.240pt}}
\multiput(896.34,192.43)(10.000,-10.426){2}{\rule{0.800pt}{0.620pt}}
\multiput(909.40,176.94)(0.512,-0.639){15}{\rule{0.123pt}{1.218pt}}
\multiput(906.34,179.47)(11.000,-11.472){2}{\rule{0.800pt}{0.609pt}}
\multiput(920.40,162.64)(0.512,-0.689){15}{\rule{0.123pt}{1.291pt}}
\multiput(917.34,165.32)(11.000,-12.321){2}{\rule{0.800pt}{0.645pt}}
\multiput(931.40,147.94)(0.512,-0.639){15}{\rule{0.123pt}{1.218pt}}
\multiput(928.34,150.47)(11.000,-11.472){2}{\rule{0.800pt}{0.609pt}}
\multiput(942.40,133.64)(0.512,-0.689){15}{\rule{0.123pt}{1.291pt}}
\multiput(939.34,136.32)(11.000,-12.321){2}{\rule{0.800pt}{0.645pt}}
\multiput(953.40,117.95)(0.526,-0.825){7}{\rule{0.127pt}{1.457pt}}
\multiput(950.34,120.98)(7.000,-7.976){2}{\rule{0.800pt}{0.729pt}}
\end{picture}

%% file: knew1.bbl
\begin{references}

\bibitem{LEGG} A. J. Leggett and A. Garg, Phys. Rev. Lett. {\bf 54}, 
857 (1985). 

\bibitem{BELL} J. S. Bell, Physics {\bf 1}, 195 (1964);
{\sl Speakable and unspeakable in quantum 
mechanics: collected papers in quantum mechanics}, 
(Cambridge University Press, Cambridge, 1987), in particular pp.14-21.

\bibitem{ASPECT} A. Aspect {\it et al.}, Phys. Rev. Lett. {\bf 47}, 460 
(1981); {\bf 49}, 91, 180 (1982).

\bibitem{PAZ} J. P. Paz and G. Mahler, Phys. Rev. Lett. {\bf 71}, 3235 
(1993).

\bibitem{SANTOS} S. F. Huelga, T. W. Marshall, and E. Santos, Phys. Rev. A
{\bf 52}, R2497 (1995).

\bibitem{CAVES} C. V. Caves, K. S. Thorne, R. W. Drever, V. Sandberg, and 
M. Zimmermann, Rev. Mod. Phys. {\bf 52}, 341 (1980).

\bibitem{BRAG} V. B. Braginsky and F. Ya. Khalili, {\sl Quantum Measurement}, 
edited by K. S. Thorne (Cambridge University Press, Cambridge, 1992), and 
references cited therein.

\bibitem{vN} J. V. von Neumann, {\sl Mathematical Foundations of Quantum 
Mechanics} (Princeton University Press, Princeton, 1955).

\bibitem{MEOP} M. B. Mensky, R. Onofrio, and C. Presilla, 
Phys. Rev. Lett. {\bf 70}, 2825 (1993).

\bibitem{REMPE} G. Rempe, H. Walther, and N. Klein, Phys. Rev. Lett. 
{\bf 58}, 353 (1987); G. Rempe, F. Schmidt-Kahler, and H. Walther, 
Phys. Rev. Lett. {\bf 64}, 2783 (1990). 

\bibitem{TESCHE} C. D. Tesche, in {\sl Proc. NY Conf. on Quantum 
Measurement Theory} (New York Academy of Sciences, New York, 1986), p. 36; 
Phys. Rev. Lett. {\bf 64}, 2358 (1990).

\bibitem{CALON} T. Calarco and R. Onofrio, Phys. Lett. A {\bf 198}, 279 
(1995).

\bibitem{ONCAL} R. Onofrio and T. Calarco, Phys. Lett. A {\bf 208}, 40 
(1995).

\bibitem{CALON1} T. Calarco and R. Onofrio, {\it Are violations to temporal 
Bell inequalities there when somebody looks?}, in preparation.
\end{references}
